\documentclass[times,doublespace]{simauth} 

\usepackage[numbers, sort&compress]{natbib}

\newtheorem{theorem}{Theorem}

\newtheorem{lemma}{Lemma}

\usepackage{booktabs}  
\usepackage{multirow}  

\newcommand\abs[1]{\left|#1\right|} 

\newcommand\BibTeX{{\rmfamily B\kern-.05em \textsc{i\kern-.025em b}\kern-.08em
T\kern-.1667em\lower.7ex\hbox{E}\kern-.125emX}}

\def\volumeyear{2019}

\begin{document}
\runninghead{Maiti \textit{et al}}
\title{A distribution-free smoothed combination method of biomarkers to improve diagnostic accuracy in multi-category classification}

\author{Raju Maiti\affil{a}\corrauth\,
	Jialiang Li\affil{a,b,c}, Priyam Das\affil{d}, Lei Feng\affil{e}, Derek Hausenloy\affil{f}, Bibhas Chakraborty\affil{a,b,g}}
\address{\affilnum{a}Centre for Quantitative Medicine, Duke-National University of Singapore Medical School, Singapore\\
	\affilnum{b}Department of Statistics and Applied Probability, National University of Singapore, Singapore\\
	\affilnum{c}Singapore Eye Research Institute, Singapore\\
	\affilnum{d}Department of Biostatistics, University of Texas MD Anderson Cancer Center, USA \\
	\affilnum{e} Department of Psychological Medicine, Yong Loo Lin School of Medicine, National University of Singapore, Singapore\\
	\affilnum{f}Cardiovascular and Metabolic Disorders Program, Duke-National University of Singapore Medical School, Singapore\\
	\affilnum{g} Department of Biostatistics and Bioinformatics, Duke University, USA}
\corraddr{}

\begin{abstract}
{Results from multiple diagnostic tests are usually combined to improve the overall diagnostic accuracy. For binary classification, maximization of the empirical estimate of the area under the receiver operating characteristic (ROC) curve is widely adopted to produce the optimal linear combination of multiple biomarkers. In the presence of large number of biomarkers, this method proves to be computationally expensive and difficult to implement since it involves maximization of a discontinuous, non-smooth function for which gradient-based methods cannot be used directly. Complexity of this problem increases when the classification problem becomes multi-category. In this article, we develop a linear combination method that maximizes a smooth approximation of the empirical Hyper-volume Under Manifolds (HUM) for multi-category outcome. We approximate HUM by replacing the indicator function with the sigmoid function or normal cumulative distribution function (CDF). With the above smooth approximations, efficient gradient-based algorithms can be employed to obtain better solution with less computing time.  We show that under some regularity conditions, the proposed method yields consistent estimates of the coefficient parameters. We also derive the asymptotic normality of the coefficient estimates. We conduct extensive simulations to examine our methods. Under different simulation scenarios, the proposed methods are compared with other existing methods and are shown to outperform them in terms of diagnostic accuracy.  The proposed method is illustrated using two real medical data sets.}
\end{abstract}


\maketitle

\section{Introduction}\label{sec1}

{Statistical classification methods are widely used in various fields such as economics, computer science, meteorology, and medicine. Specifically, in medicine, diagnostic tests are employed as effective ``classifiers" to discriminate diseased individuals from the non-diseased. Over the recent decades, many research articles recommended combining multiple test results in order to increase the overall diagnostic accuracy. Common approaches to combine multiple test results include the logistic regression (LR), the linear discriminant analysis (LDA) and other model-based approaches. Some authors (\cite{su1993liu}, \cite{pepe2000thompson}, \cite{pepe2006cai}) directly focused on the maximization of the Area Under the Receiver Operating Characteristic (ROC) Curve (AUC) to combine multiple test results. However, to the best of our knowledge, there is only limited development for finding the optimal linear combination of diagnostic tests in case of multivariate classification problems.}

{For binary classification, earlier works considered maximizing various non-parametric estimates of AUC to obtain the best linear combination of the biomarkers (\cite{pepe2000thompson}, \cite{ma2005}, \cite{ma2006}, \cite{ma2007huang}, \cite{liu2011min}, among others). In particular, \cite{pepe2006cai} proposed to maximize an empirical estimate of AUC in the form of a Mann-Whitney U-statistic for obtaining the best solution. However, maximization of the empirical AUC remains computationally challenging since the objective function is discontinuous and non-differentiable. To reduce the computational complexity, \cite{ma2007huang} considered maximizing a smooth consistent approximation of the empirical AUC using the sigmoid function to estimate the optimal coefficient parameters for the binary classification scenario. For multivariate classification problems, \cite{liu2011min} proposed a min-max method where only the biomarkers with the minimum and the maximum values are considered for each subject, and then they are combined linearly by maximizing the empirical AUC. Thus, irrespective of the number of biomarkers, min-max method estimates only one coefficient at a time which is computationally less expensive.}

When a disease outcome involves more than two categories, Hyper-volume Under the ROC Manifold (HUM) is commonly used as a multi-category extension of AUC (\cite{li2008fine}). In such problems, the goal is to find the optimal combination of biomarkers that maximizes the diagnostic accuracy measure HUM. For a three-category outcome, HUM is also known as the Volume under ROC Surface (VUS), and has also been considered in the context of some real applications (\cite{scurfield1996}, \cite{mossman1999}, \cite{nakas2004yiannoutsos}). To evaluate the HUM values for single marker or multiple markers under existing learning methods, one may adopt R packages {\tt HUM} \cite{novo2013} and {\tt mcca} \cite{li2019}, respectively. \cite{zhang2011li} maximized the empirical estimate of VUS to combine multiple biomarkers. Due to non-differentiability of the objective function, maximization of empirical VUS requires derivative-free optimization methods which are computationally expensive, especially when the number of biomarkers is large. To overcome this problem, under normality assumption, \cite{kang2013} used a penalized and scaled stochastic distance method to combine multiple biomarkers, which was computationally less demanding. However, violation of the normality assumption of biomarkers may lead to poor estimation performance. \cite{hsu2016chen} constructed upper and lower bounds of the HUM using Fr\'echet inequality and showed that these bounds are functions of AUCs of all possible pairwise adjacent categories. Then they maximized the empirical estimates of such upper and lower bounds to obtain the optimal linear combination. This technique reduces the computational complexity. However, such approximations do not perform well for small sample sizes and/or non-normal distributions (as is observed in our simulation study).

{In this article, we propose to maximize the distribution-free Smooth approximation of empirical HUM (SHUM) to combine multiple biomarkers in an effective way. In particular, the sigmoid function and the normal cumulative distribution functions (CDF) are used to approximate the non-differentiable indicator functions embedded in the definition of HUM. We show that the proposed method yields consistent estimates of the optimal coefficients and they are asymptotically normal. A major advantage with the proposed method stems from the fact that SHUM is a continuous and differentiable function; this feature of SHUM allows one to adopt a variety of gradient-based optimization algorithms. Maximizing empirical HUM with derivative-free optimization techniques, such as Nelder-Mead simplex method, genetic algorithm (GA), and simulating annealing (SA), are computationally expensive. However, gradient-based optimization techniques like Newton-Raphson and Quasi-Newton methods can be applied to maximize the SHUM function; these nonlinear solvers are much more stable with nice convergence properties. In addition to the theoretical developments, we also carry out extensive simulations to examine our methods and compare their performance with other existing methods, e.g., the min-max method (\cite{liu2011min}), the lower and upper bound methods (\cite{hsu2016chen}), the empirical method (\cite{zhang2011li}) and the parametric method with normal distribution (\cite{zhang2010thesis}).}

{As a motivating application, we consider data from the {\bf E}ffect of {\bf R}emote {\bf I}schemic Preconditioning on
{\bf C}linical Outcomes in Patient Undergoing {\bf C}oronary {\bf A}rtery Bypass Graft Surgery (ERICCA) trial where a group of patients participated in a cardiovascular surgery and were followed for one year after the surgery (\cite{hausenloy2015}). 
During the study period, patients might have developed Acute Kidney Injury (AKI) which was recorded as a multi-category ordinal outcome with 4 severity levels. In another application, we consider data on Alzheimer's disease from the Alzheimer’s Disease Research Center (ADRC) at the  University of Washington. There, based on the level of disease severity, the patients were divided into 3 groups and data on 14 biomarkers were collected. For both the datasets, we apply our proposed methods to combine the biomarkers and compare the results with the competing methods.}

The rest of the article is organized as follows. In Section 2, HUM and SHUM are defined along with discussion on the large sample properties of the estimated combination coefficients. In Section 3, existing methods are summarized in an overview. In Section 4, we provide a discussion on computational issues. In section 5, we present results from the simulation studies. Section 6 describes the results and findings from two real data analyses. Section 7 contains discussion and concluding remarks. All the proofs of theoretical results can be found in the Appendix.

\section{Methods}

In this section, we introduce the  HUM and SHUM methods for combining multiple markers to improve the multi-category classification accuracy.

\subsection{{Hyper-volume Under ROC Manifold (HUM)}}

Consider a study where there are $M$ multiple diagnostic/disease categories which are assumed to be ordered in nature without loss of generality. We provide some practical suggestion later for unordered classes. Suppose  
$\mathbf{X}_{1}, \mathbf{X}_{2}, \cdots, \mathbf{X}_{M}$ are $d$-dimensional random selected vectors representing the values of $d$ biomarkers for $M$ diagnostic/disease categories where $\mathbf{X}_{j} = (X_{j1}, X_{j2},\cdots, X_{jd})^{T}$ and $X_{jk}$ denotes the value of the $k$-th biomarker from the $j$-th category, $k=1,2,\cdots,d$ and $j=1,2,\cdots,M$. Suppose $\mathbf{X}_{j}$ follows  multivariate continuous distribution $F_{j}$. Consider a linear combination of these biomarkers as $$ \boldsymbol{\beta}^{T}\mathbf{X}_{j}=\displaystyle\sum_{k=1}^{d}\beta_{j}X_{jk}, \; j=1,2,\cdots, M, $$ 
where $\boldsymbol{\beta}=(\beta_{1}, \beta_{2}, \cdots, \beta_{d})^{T}$ is a $d$-dimensional vector of parameters. Under the assumption that the larger value of the above combination corresponds to more severe disease category, a diagnostic accuracy measure can be defined by the following probability
$$D(\boldsymbol{\beta})  = P(\boldsymbol{\beta}^{T}\mathbf{X}_{M}>\boldsymbol{\beta}^{T}\mathbf{X}_{(M-1)}>\cdots>\boldsymbol{\beta}^{T}\mathbf{X}_{1}),$$
which is known as hyper-volume under the ROC manifolds (HUM) (\cite{scurfield1996}, \cite{li2008fine}). For multi-category ordinal outcome, HUM can be considered as an extension of the AUC which is widely used for binary diagnostic accuracy studies. Our objective is to find the best possible value of $\beta$ for which $D(\boldsymbol{\beta})$ is maximized. Ideally, if there exist a $\beta$ for which $D(\boldsymbol{\beta}) = 1$, using such a combination the diagnostic categories would be perfectly separated. Let $\boldsymbol{\beta}_{0}$ denote the optimal coefficient parameter that maximizes $D(\boldsymbol{\beta})$ over a restricted parametric space $B$, 
$$\boldsymbol{\beta}_{0} = \arg \max_{\boldsymbol{\beta} \in B} D(\boldsymbol{\beta}),$$ 
where the restricted space $B = \{\boldsymbol{\beta}\in \mathbb{R}^{d}: \beta_{d} = 1\}$ is considered to avoid the identifiability problem. Denote $\boldsymbol{\theta} = (\beta_{1}, \beta_{2}, \cdots, \beta_{d-1})^T$ to be the first $d-1$ components of $\boldsymbol{\beta}$ which are the actual coefficient parameters free to vary in the $d-1$ dimensional Euclidean space.  Hereafter, for the simplicity of presentation, we use  $\boldsymbol{\beta}$ in place of $\boldsymbol{\beta}(\boldsymbol{\theta})=(\boldsymbol{\theta}^{T}, 1)^T$. If the biomarkers are non-informative in predicting the disease categories then $D(\boldsymbol{\beta})$ will be close to the probability of a random sorting $\frac{1}{M!}$. Under the assumption that $\mathbf{X}_{1}, \mathbf{X}_{2}, \cdots, \mathbf{X}_{M}$ are generated from multivariate normal distribution, a unique solution for $\boldsymbol{\beta}_{0}$ can be derived  under some regularity conditions, (\cite{su1993liu}). 
However, in general for non-normal data, there does not exist such closed form expression of $\boldsymbol{\beta}_{0}$ and numerical optimizer must be utilized.

\subsection{{Empirical HUM}}\label{sec_EHUM}
{Now let us consider the problem of estimating $\boldsymbol{\beta}_{0}$  given an empirical sample. Let $\{\mathbf{X}_{ji_j}; \; i_j=1,2,\cdots,n_{j}, \; j=1,2,\cdots,M\}$  be a sample of size $n=\sum_{j=1}^{M}n_{i}$ observations where $j=1,\ldots,M$ denote diagnostic categories and $i_{j}=1,2,\cdots,n_{j}$ denote the samples  in the $j$-th category. Then, for a fixed $\boldsymbol{\beta}$, the empirical HUM  is  given by
\begin{eqnarray}\label{eqn-hum-empirical}
 {D}_{E}(\boldsymbol{\beta}) &=& \dfrac{1}{n_{1}n_{2} \cdots n_{M}} \displaystyle\sum_{i_{1}=1}^{n_{1}}\displaystyle\sum_{i_{2}=1}^{n_{2}} \cdots \displaystyle\sum_{i_{M}=1}^{n_{M}}I(\boldsymbol{\beta}^{T}\mathbf{X}_{Mi_{M}}>\boldsymbol{\beta}^{T}\mathbf{X}_{(M-1)i_{(M-1)}}>\cdots>\boldsymbol{\beta}^{T}\mathbf{X}_{1i_{1}})  \nonumber \\
  &=& \dfrac{1}{\displaystyle\prod_{i=1}^{M}n_{i}} \displaystyle\sum_{i_{1}=1}^{n_{1}}\displaystyle\sum_{i_{2}=1}^{n_{2}} \cdots \displaystyle\sum_{i_{M}=1}^{n_{M}} I(\boldsymbol{\beta}^{T}\mathbf{X}_{Mi_{M}}>\boldsymbol{\beta}^{T}\mathbf{X}_{(M-1)i_{(M-1)}}) \cdots I(\boldsymbol{\beta}^{T}\mathbf{X}_{2i_{2}}>\boldsymbol{\beta}^{T}\mathbf{X}_{1i_{1}})  \nonumber \\
\end{eqnarray}
where $I(\cdot)$ denotes the indicator function. When sample size is large, $D_E(\boldsymbol{\beta})$ is a very close approximation to $D(\boldsymbol{\beta})$. Therefore an optimal coefficient parameter can be estimated by
$$ \widehat{\boldsymbol{\beta}}_{E} = \arg \max_{\boldsymbol{\beta} \in B} {D}_{E}(\boldsymbol{\beta}).$$
When the number of disease categories is 2 (i.e., $M=2$), the empirical HUM reduces to the empirical estimate of AUC given by $$D_{E}(\boldsymbol{\beta}) =  \dfrac{1}{n_{1}n_{2}} \displaystyle\sum_{i_{1}=1}^{n_{1}}\displaystyle\sum_{i_{2}=1}^{n_{2}} I(\boldsymbol{\beta}^{T}\mathbf{X}_{2i_{2}}>\boldsymbol{\beta}^{T}\mathbf{X}_{1i_{1}}),$$
and when $M=3$, it reduces to the empirical VUS given by 
{\begin{align}
D_{E}(\boldsymbol{\beta}) =  \dfrac{1}{n_{1}n_{2}n_{3}} \displaystyle\sum_{i_{1}=1}^{n_{1}}\displaystyle\sum_{i_{2}=1}^{n_{2}}\displaystyle\sum_{i_{3}=1}^{n_{3}} I(\boldsymbol{\beta}^{T}\mathbf{X}_{3i_{3}}>\boldsymbol{\beta}^{T}\mathbf{X}_{2i_{2}}>\boldsymbol{\beta}^{T}\mathbf{X}_{1i_{1}}).
\label{EVUS}
\end{align}}

Under some regularity conditions, \cite{zhang2011li} established the consistency and asymptotic normality of $\widehat{\boldsymbol{\beta}}_{E}$ for three-category outcome. Following their argument, the consistency and asymptotic normality of $\widehat{\boldsymbol{\beta}}_{E}$ for more than three categories can be similarly established. However, upon close examination, we notice that ${D}_{E}(\boldsymbol{\beta})$ is discontinuous and not differentiable w.r.t. $\boldsymbol{\beta}$, and hence faster gradient-based algorithms are not useful to this optimization problem. On the other hand, although derivative-free algorithms can be used for small number of categories, say $M=2$ or $3$, with the increase in the number of categories derivative-free algorithms become computationally prohibitive and instable. To address this issue, in the next section, we propose a new method based on smooth approximation.}

\subsection{{Smooth Approximation of empirical HUM}}
In order to alleviate the computational burden of maximizing the sample version HUM, as an alternative we propose to maximize a class of smooth approximations of the empirical HUM. The basic idea is to approximate the non-differentiable indicator function $I(x>0)$. We focus on the class of all continuous distribution functions $g(x)$ with support over $(-\infty,\infty)$, satisfying $g(x)+g(-x)=1$ and  $g^{''}(x)$ is continuous. Having $g^{''}(x)$ continuous and replacing all indicator functions with this kind of approximation function makes the approximate objective function solvable with gradient-based optimization algorithms such as the Newton-Raphson methods and the Quasi-Newton methods. In this paper, we consider two smooth candidates which are the sigmoid function $s(x) = \frac{1}{1+\exp(-x)}$, and the standard normal CDF denoted by $\Phi(x)=P(\chi \leq x)$ where $\chi$ follows a normal distribution with mean 0 and variance 1. Under the binary classification scenario, \cite{ma2007huang} proposed the sigmoid approximation of the empirical AUC to seek $\boldsymbol{\beta}_{0}$. However this approach has never been extended for multi-category classification scenario to the best of our knowledge. 

As the value of $x$ goes away from 0, $s(x)$ tends to get closer to $I(x)$. When $x$ is close to 0, the absolute difference between $s(x)$ and $I(x)$ is the highest. This also holds true for $\Phi(x)$. Therefore, in order to improve the approximation of these functions, a tuning parameter $\lambda_{n}$ is introduced and we approximate $I(x)$ by 
   $s_{n}(x)=s(\frac{x}{\lambda_{n}})=\frac{1}{1+\exp(-x/\lambda_{n})}$ and    $\Phi_{n}(x) = \Phi(x/\lambda_{n})$ where $\lambda_{n}$ satisfies $\lim_{n \rightarrow \infty} \lambda_{n} = 0$.
   
\begin{figure}[t]
 \centering
  \includegraphics[scale=0.6]{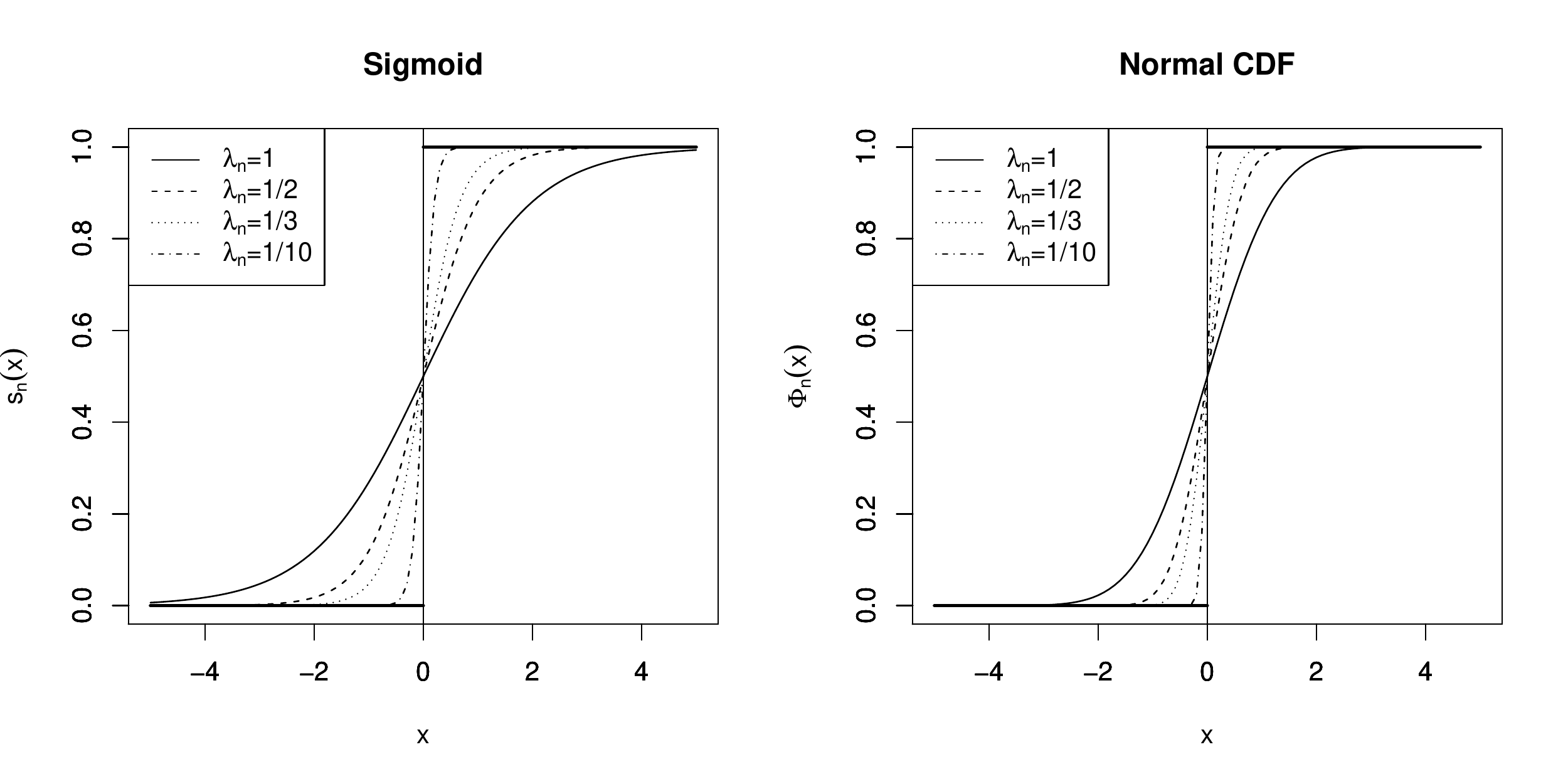}
 \caption{Sigmoid and normal CDF functions for different choices of tuning parameter $\lambda_n$}
 \label{fig-lambda-n}
\end{figure}

 The choice of $\lambda_{n}$ is very crucial in the performance of the smoothed HUM function. When $\lambda_{n}$ is close to 0, the proposed SHUM estimator behaves similarly to the empirical HUM with a very large value of derivative across a very small interval around zero. This induces a greater variability on the resulting estimators. On the other hand, if $\lambda_{n}$ is chosen to be one, it suffers from biased approximation.  Therefore, we need to choose an optimal $\lambda_{n}$ between 0 and 1 to strike a balance between the bias and the variance issues.  To illustrate the role of $\lambda_{n}$, a graphical representation is displayed in Figure \ref{fig-lambda-n} where we consider a few selected values of $\lambda_n$. We can see that as $\lambda_{n}$ decreases to zero the approximation becomes closer to the indicator function $I(x)$. As a rule of thumb, \cite{gammerman1996} and \cite{ma2007huang} recommended $\lambda_{n}$ should be chosen ensuring that $|\boldsymbol{\beta}^{T}(\mathbf{X}_{1i_1} - \mathbf{X}_{2i_2})/ \lambda_{n}| > 5$ is satisfied for most of the pairs  ($i_1$, $i_2$). In this paper after experimenting with different possible values of $\lambda_{n}$, we set $\lambda_{n}=\frac{1}{\sqrt{n}}$ for our simulation studies and real data analysis, which satisfies the empirical condition.

{Although the smoothing approximation can be done through either $g_{n}=s_n$ or $g_n=\Phi_n$, hereafter we only present the results for the sigmoid smoothing approximation to save the space. Applying this proposed function approximation to ${D}_{E}(\boldsymbol{\beta})$, the proposed sigmoid smooth approximation function for multi-categorical problem is given by
\begin{equation}\label{eqn-hum-sigmoid}
  {D}_{s_{n}}(\boldsymbol{\beta}) = \dfrac{1}{n_{1}n_{2}\cdots n_{M}} \displaystyle\sum_{i_{1}=1}^{n_{1}}\displaystyle\sum_{i_{2}=1}^{n_{2}} \cdots \displaystyle\sum_{i_{M}=1}^{n_{M}} s_{n}(\boldsymbol{\beta}^{T}(\mathbf{X}_{Mi_{M}}-\mathbf{X}_{(M-1)i_{(M-1)}})) \cdots s_{n}(\boldsymbol{\beta}^{T}(\mathbf{X}_{2i_{2}}-\mathbf{X}_{1i_{1}})).
\end{equation}
We propose to maximize ${D}_{s_{n}}(\boldsymbol{\beta})$ in order to estimate the optimal coefficient vector. The optimal vector of combination is estimated by
$$\widehat{\boldsymbol{\beta}}_{s_{n}} = \arg \max_{\boldsymbol{\beta} \in B} {D}_{s_{n}}(\boldsymbol{\beta}).$$
We denote the optimal coefficient estimate obtained using the sigmoid smooth approximation of the empirical HUM (SSHUM) as $\widehat{\boldsymbol{\beta}}_{s_{n}}$ and using the normal smooth approximation of the empirical HUM (NSHUM) by $\widehat{\boldsymbol{\beta}}_{\Phi_{n}}$.


\subsection{Consistency and Asymptotic Normality of SSHUM}
Under some regularity conditions, we establish the consistency and asymptotic normality of $\widehat{\boldsymbol{\beta}}_{s_{n}}$. We list the set of necessary regularity conditions as follows.
\begin{itemize}
 \item[A1.] The support space of $\mathbf{X}_{ji_j}$ is not contained in any proper linear subspace of $\mathbb{R}^{d}$.
 \item[A2.] There exist at least one component of $\mathbf{X}_{ji_j}$ which has positive density everywhere conditional on the other components, almost surely.
 \item[A3.] The true parameter value $\boldsymbol{\beta}_{0}$ is an interior point of $B$ which is a compact subset of $\mathbb{R}^{d}$.
\end{itemize}

\begin{theorem}[Consistency]\label{theo_cons}
 Suppose that assumptions (A1)-(A3) hold, then
 $$\widehat{\boldsymbol{\beta}}_{s_n} \overset{p}{\longrightarrow} \boldsymbol{\beta}_{0}$$ as $n \rightarrow \infty$, where $``{\overset{p}{\longrightarrow}}"$ denotes {\it convergence in probability}.
\end{theorem} 
The detailed proof of Theorem \ref{theo_cons} is provided in section A1 in the appendix. {In order to prove the asymptotic normality of $\widehat{\boldsymbol{\beta}}_{s_n}$, we assume additional set of regularity conditions. Denote $\Psi(\mathbf{X}_{1i_1}, \mathbf{X}_{2i_2}, \mathbf{X}_{Mi_M}; \boldsymbol{\beta})=\dfrac{\partial}{\partial\boldsymbol{\theta}}  \left[s_{n}(\boldsymbol{\beta}^{T}(\mathbf{X}_{Mi_{M}}-\mathbf{X}_{(M-1)i_{(M-1)}})) \cdots s_{n}(\boldsymbol{\beta}^{T}(\mathbf{X}_{2i_{2}}-\mathbf{X}_{1i_{1}})) \right]$. Then assume the following:
\begin{itemize}
 \item[A4.] $\mathbf{A}(\boldsymbol{\beta}_{0}) = E\left(-\dfrac{\partial}{\partial\boldsymbol{\theta}^{T}}\Psi(\mathbf{X}_{1i_1}, \mathbf{X}_{2i_2}, \mathbf{X}_{Mi_M}; \boldsymbol{\beta}_{0})\right)<\infty$  and is invertible.
 \item[A5.] $\tilde{\Psi}_{m1}(\mathbf{X}_{m1}; \boldsymbol{\beta}_{0}) = E\left(\dfrac{\partial}{\partial\boldsymbol{\theta}}\Psi(\mathbf{X}_{11}, \mathbf{X}_{21}, \mathbf{X}_{M1}; \boldsymbol{\beta}_{0})|\mathbf{X}_{m1}\right)$ is having the finite variance-covariance matrix, i.e., $\Sigma_{\psi_{m}} = Var(\tilde{\Psi}_{m1}(\mathbf{X}_{m1}; \boldsymbol{\beta}_{0})) <\infty$ for all $m=1,2,\cdots,M$.
 \item[A6.] $\lim_{n \rightarrow \infty}\dfrac{n}{n_{m}}=\rho_{m}^{2} < \infty$ for all $1\le m \le M$. 
\end{itemize}
Assumptions (A4)-(A6) ensure that the asymptotic variance exits and is finite.}

\begin{theorem}[Asymptotic normality]\label{theo_normality}
 Suppose that the regularity conditions (A1)-(A6) hold,  then 
 $$\sqrt{n}(\widehat{\boldsymbol{\beta}}_{s_n} - \boldsymbol{\beta}_{0}) \overset{D}{\longrightarrow} (W^{T},0)^{T}$$
 as $n \longrightarrow \infty$ where $``{\overset{D}{\longrightarrow}}"$ denotes {\it convergence in distribution} and $W$ is a $(d-1)$-variate normal distribution $N(\mathbf{0}, \mathbf{A}^{-1}(\boldsymbol{\beta}_{0})\mathbf{B}(\boldsymbol{\beta}_{0})\{\mathbf{A}^{-1}(\boldsymbol{\beta}_{0})\}^{T})$, 
 where 
 \begin{eqnarray*}
   \mathbf{B}(\boldsymbol{\beta}_{0}) &=& \displaystyle\sum_{m=1}^{M}\rho_{m}^{2}\Sigma_{\psi_{m}}.
 \end{eqnarray*}
 \end{theorem}

{Remark: Computation of variance of $\widehat{\boldsymbol{\beta}}_{s_n}$ using the asymptotic variance formula given in Theorem 2 is very tedious and challenging, especially because of the choice of the kernel function $\mathbf{g}$ given in equation \eqref{eqn-g}. Furthermore, it is noticed that the U-statistic based asymptotic variance formula are not generally reliable for small sample size for the direct maximization of the empirical HUM (see \cite{li2008fine}). In such cases, bootstrap technique is usually employed to compute the variances of the coefficient estimators of ${\boldsymbol{\beta}}_{s_n}$.} 


\section{Existing Methods}\label{sec-existing}
{In this section, we provide a brief summary of the existing methods which can be used to obtain the optimal coefficient vector for biomarker combinations. In the simulation study section, we shall compare the proposed methods with these methods.}



\subsection{Parametric Method with Normality Assumption (Parametric)}
\cite {zhang2010thesis} proposed the parametric method with normality assumption of the biomarkers in order to obtain the optimal linear combination of biomarkers. This approach assumes that $F_j$, the distribution of biomarkers from the $j$th category $\boldsymbol{X}_{j} $, is a multivariate normal distribution with mean vector $\boldsymbol{\mu}_{j}$ and variance-covariance matrix $\Sigma_{j}$, $j=1,2,\cdots,M$. Then, the linear combination of biomarkers $\boldsymbol{X}_{j}$ for the $j$-th category, denoted by $V_{j} = \boldsymbol{\beta}^{T}\boldsymbol{X}_{j}$, follows a univariate normal distribution with mean $\boldsymbol{\beta}^{T}\boldsymbol{\mu}_{j}$ and variance $\boldsymbol{\beta}^{T}\Sigma_{j}\boldsymbol{\beta}$, i.e., $V_{j} \sim N(\boldsymbol{\beta}^{T}\boldsymbol{\mu}_{j}, \boldsymbol{\beta}^{T}\Sigma_{j}\boldsymbol{\beta})$ $j=1,2,\cdots,M$. Let $\phi$ and $\Phi$ denote the density function and cumulative distribution function of the standard normal distribution $N(0,1)$. For $M=3$, the HUM $D(\boldsymbol{\beta})$ can be shown to be equal to 
\begin{eqnarray}
 D_{N}(\boldsymbol{\beta}) &=& \int_{-\infty}^{\infty} \Phi\left(\dfrac{\sqrt{\boldsymbol{\beta}^{T}\Sigma_{2}\boldsymbol{\beta}}}{\sqrt{\boldsymbol{\beta}^{T}\Sigma_{1}\boldsymbol{\beta}}} u + \dfrac{\boldsymbol{\beta}^{T}(\boldsymbol{\mu}_{2} - \boldsymbol{\mu}_{1})}{\sqrt{\boldsymbol{\beta}^{T}\Sigma_{1}\boldsymbol{\beta}}} \right)  \Phi\left(- \dfrac{\sqrt{\boldsymbol{\beta}^{T}\Sigma_{2}\boldsymbol{\beta}}}{\sqrt{\boldsymbol{\beta}^{T}\Sigma_{3}\boldsymbol{\beta}}} u + \dfrac{\boldsymbol{\beta}^{T}(\boldsymbol{\mu}_{3} - \boldsymbol{\mu}_{2})}{\sqrt{\boldsymbol{\beta}^{T}\Sigma_{3}\boldsymbol{\beta}}} \right) \phi(u) \; du. \nonumber \\ 
 \end{eqnarray}\label{eqn-normal-hum}
 Maximizing $D_{N}(\boldsymbol{\beta})$  with respect to $\boldsymbol{\beta}$, we obtain the optimal coefficient estimates as 
 $$\boldsymbol{\beta}_{N} = \arg \max_{\boldsymbol{\beta} \in B} D_{N}(\boldsymbol{\beta}).$$
Following the results of \cite{su1993liu}, it can be shown that if $\mathbf{X}_{1}, \mathbf{X}_{2}, \cdots, \mathbf{X}_{M}$ are  multivariate normally distributed with mean vectors $\boldsymbol{\mu}_{1}$, $\boldsymbol{\mu}_{2}$, $\cdots$, $\boldsymbol{\mu}_{M}$, respectively and common variance-covariance matrix $\Sigma$ satisfying 
\begin{eqnarray}\label{eqn-normal-cond}
\boldsymbol{\mu}_{2}-\boldsymbol{\mu}_{1}=\boldsymbol{\mu}_{3}-\boldsymbol{\mu}_{2}= \cdots = \boldsymbol{\mu}_{M}-\boldsymbol{\mu}_{M-1} =\boldsymbol{\delta}, 
\end{eqnarray}
the optimal coefficient parameters $\widehat{\boldsymbol{\beta}}_{N}$ will be proportional to $\Sigma^{-1}\boldsymbol{\delta}$, i.e., $\widehat{\boldsymbol{\beta}}_{N}  \propto \Sigma^{-1}\boldsymbol{\delta}$. Once we have the sample estimates for the mean and covariance parameters, we can plug-in them into the formula of $\widehat{\boldsymbol{\beta}}_{N}$ to obtain the coefficient estimates. 

{A major advantage of using normality assumption is that it is computationally very easy, especially when \eqref{eqn-normal-cond} holds true. However, the method fully depends on the normality assumption. Violation of the normality assumption may result in poor estimate of $\boldsymbol{\beta}_{0}$.}

\subsection{Min-Max Method (Min-Max)}
The Min-Max {(MM)} method is a more simplified non-parametric approach to combine the multiple biomarkers. It was originally proposed by \cite{liu2011min} in the context of binary outcome. Instead of considering all the biomarkers, this method considers the empirical AUC based on the linear combination of two extreme biomarkers for each subject in the study. In this paper, to facilitate a comparative study, we define the empirical HUM based on the combination of the minimum and maximum biomarkers for each subject. 

{Let $X_{ji_j,max} = \max_{1\le k \le d} X_{ji_j,k}$ and  $X_{ji_j,min} = \min_{1\le k \le d} X_{ji_j, k}$ and define the linear combination of  these two extreme observations as $ V_{ji_j} = \beta_{max} X_{ji_j,max} + \beta_{min}  X_{ji_j,min}$, $i=1,2,\cdots,n_j$, $j=1,2,\cdots,M$. Then the objective function to be maximized to obtain the optimal coefficient vector is given by
\begin{eqnarray}\label{eqn-minmax-hum}
 D_{MM}(\boldsymbol{\beta}) &=& \dfrac{1}{\displaystyle\prod_{j=1}^{M}n_{j}} \displaystyle\sum_{i_{1}=1}^{n_{1}}\displaystyle\sum_{i_{2}=1}^{n_{2}} \cdots \displaystyle\sum_{i_{M}=1}^{n_{M}}I(V_{Mi_M} > V_{(M-1)i_{M-1}} > \cdots > V_{1i_{1}}).
\end{eqnarray}
The optimal coefficient estimates by maximizing the above quantity can be written as
$$ \widehat{\boldsymbol{\beta}}_{MM} = \arg \max_{\boldsymbol{\beta}\in B} {D}_{MM}(\boldsymbol{\beta}).$$}

{A major advantage of this method is that it involves the optimization of a single parameter as opposed to other competing methods, and hence computationally it is very efficient. Furthermore, it does not need to assume any distributional assumption of the data and hence is more robust against the parametric methods. So far, the method is studied only in the context of binary disease outcome and it is observed that the method can achieve higher sensitivity over a certain range of specificity. In other words, when someone is interested in partial AUC, this methods works better. However a major limitation of this method is that a major portion of the informations on the biomarkers are not utilized since only maximum and minimum biomarkers’ values are used.}

\subsection{{Upper and Lower Bound Approach using Fr\'echet inequality (Fr\'echet)} }
To reduce computational burden of maximizing the empirical HUM in case of higher number of disease categories and/or number of biomarkers, \cite{hsu2016chen} proposed the upper and lower bounds of HUM which are given by
$$\max\{0, (M-1)P_{A}(\boldsymbol{\beta}) - (M-2)\} \le D(\boldsymbol{\beta}) \le P_{M}(\boldsymbol{\beta}),$$
where $P_{A}(\boldsymbol{\beta})$ and $P_{M}(\boldsymbol{\beta})$ are defined as follows
$$P_{A}(\boldsymbol{\beta}) = \displaystyle\sum_{j=1}^{M-1}P(\boldsymbol{\beta}^{T}\mathbf{X}_{j+1}>\boldsymbol{\beta}^{T}\mathbf{X}_{j})/(M-1),$$
and $$ P_{M}(\boldsymbol{\beta}) = \min_{1 \le j \le M-1} P(\boldsymbol{\beta}^{T}\mathbf{X}_{j+1}>\boldsymbol{\beta}^{T}\mathbf{X}_{j}).$$
Instead of maximizing HUM, they proposed to maximize $P_{A}(\boldsymbol{\beta})$ or $P_{M}(\boldsymbol{\beta})$ in order to obtain the optimal combination. For example, maximizing $P_{M}(\boldsymbol{\beta})$ with respect to $\boldsymbol{\beta}$ we obtain $\widehat{\boldsymbol{\beta}}_{Frechet} = \arg \max_{\boldsymbol{\beta} \in B} P_{M}(\boldsymbol{\beta})$ which can be considered as an optimal coefficient vector.

The above method is computationally efficient against the direct maximization of HUM as it only considers pairs from the adjacent categories, i.e., binary outcomes. The above method is computationally less time consuming than the HUM when the number of 
disease categories is more than two. However, when pairwise discrimination among the disease categories are not relevant to the overall discrimination, this method might perform poorly.

\section{Step-down Algorithm for Optimization}
Step-down algorithm was originally proposed by \cite{pepe2006cai} to combine multiple biomarkers in the context of binary diagnostic outcomes. {The main motivation of using step-down algorithm is its ability to optimize the elements of the $\boldsymbol{\beta}$ vector sequentially one at a time instead of attempting to optimize them simultaneously. \cite{kang2013} formalized the step-down algorithm in the context of three-category diagnostic outcomes. Recently \cite{hsu2016chen} used this algorithm to maximize upper or lower bound of HUM and obtained an optimal linear coefficient estimates. The algorithm to maximize a criteria function (e.g., SHUM) goes as follows:}
\begin{itemize}
 \item[{\bf Step 1.}] Compute the SHUM for each individual $d$ biomarkers {using one at a time} and {arrange covariates in decreasing order} with respect to the computed SHUM values {such that $X_{(1)}$ and $X_{(d)}$ have the highest and the lowest individual SHUM values respectively.}.
 \item[{\bf Step 2.}] {Choose the first two biomarkers with the highest SHUM values and combine them as $V_2=X_{(1)} + \lambda_2 X_{(2)}$.}
\item[{\bf Step 3.}] {Maximize the SHUM for the combined marker $V_2$ w.r.t. $\lambda_2$ and obtain $\widehat{V}_2 = X_{(1)} + \widehat{\lambda}_2 X_{(2)}$.}
 \item[{\bf Step 4.}] {For $i = 3,\ldots,d$ construct $V_i = \widehat{V}_{i-1}+\lambda_i X_{(i)}$ and maximize $V_i$ w.r.t. $\lambda_i$ and obtain  $\widehat{\lambda}_i$.} 
\end{itemize}
Thus at the end of step 4, the estimated optimal combination $\widehat{V}_d = X_{(1)} + \widehat{\lambda}_2 X_{(2)} + \cdots + \widehat{\lambda}_d X_{(d)}$ is obtained. Although this algorithm has been widely used to maximize empirical HUM for binary and three-category cases, here we mainly use a gradient-decent based algorithm, namely quasi-Newton method to maximize all the stepwise SHUM values. We implement the numerical method using the in-built function {\tt optim} in the {\tt R} software freely available in {\tt www.cran.org}.

\section{Simulation Study}

To compare the performance of the proposed method with the existing methods, we perform experiments based on various simulation scenarios. We consider three biomarkers and three-category ordinal outcome $Y \in \{0, 1, 2\}$, such that higher values of biomarkers represent higher disease category. To explore the performance of the methods under different case scenarios, we consider three examples based on normal distribution (with different correlation structure) and one based on Weibull distribution to represent the non-normal and skewed family.

{\bf Scenario 1 :} For the $i$-th category, the values of the biomarkers are simulated from three dimensional normal distributions with mean vector $\boldsymbol{\mu}_{i}$, and common variance covariance matrix as identity $\Sigma=\mathbf{I}$; $i=0,1,2$. We set the parameter values as $\boldsymbol{\mu}_{0}=(0, 0, 0)^{T}$,  $\boldsymbol{\mu}_{1}=(1.0, 1.1, 1.2)^{T}$, and $\boldsymbol{\mu}_{2}=(2.0, 2.2, 2.4)^{T}$ for categories $i=0,1,2$, respectively. Since the correlation matrix is considered to be identity with normal distributions, the biomarkers are independent to each other.

{\bf Scenario 2 :} In the second scenario, the mean vectors are same as in Scenario 1, however the covariance matrix $\Sigma=((\sigma_{st}))$ is such that all the diagonal elements are 1, i.e, $\sigma_{ss}=1$; and all the off-diagonal elements are 0.2, i.e., $\sigma_{st} = 0.2, s\ne t$; $s,t=1,2,3$. This variance covariance matrix is an example of exchangeable matrix. Since all the off-diagonal elements are non-zero and equal, therefore the biomarkers are correlated.

{\bf Scenario 3 :} In the third scenario, the mean vectors are same as the previous scenarios. The covariance matrix has an AR(1) form, i.e., all the diagonal elements are 1; and the off-diagonal elements are set as $\sigma_{st} = 0.2^{|s-t|},\; s\ne t$; $s,t=1,2,3$. Here all the mutual correlations are non-zero but it fades as the distance between two biomarkers increases.

{\bf Scenario 4 :} In the fourth scenario, values of the biomarkers are simulated from Weibull distribution. Specifically, the $j$-th biomarker from the $i$-th disease category follows a Weibull distribution with shape parameter $k_j$ and scale parameter $\lambda_i$ and the probability density function is given by  
\begin{eqnarray*}
 f(x; k_j, \lambda_i) &=& \begin{cases}
                       \dfrac{k_i}{\lambda_i} \left(\dfrac{x}{\lambda_i}\right)^{k_j-1} \exp(-(\frac{x}{\lambda_j})^{k_j}) & \qquad x>0, \\
                       0 & \qquad x \le 0,
                      \end{cases} 
\end{eqnarray*}
$i=0,1,2$ and $j=1,2, 3$. Values of the shape parameter $k$ and scale parameter $\lambda$ are set as $(k_1, k_2, k_3)=(0.5,1,1.5)$ and $(\lambda_1, \lambda_2, \lambda_3)=(1,2,3)$, respectively.  Here, we assume that biomarkers are independently distributed. This case corresponds to non-normal and skewed distribution.

\subsection{Performance Evaluation}

For each of the above-mentioned scenarios, we considered three sample sizes $n=60, 90, 120$. Performance of the proposed SSHUM and NSHUM methods are compared with the existing methods, namely the empirical method (\cite{zhang2011li}), the Frechet bounds method (\cite{hsu2016chen}), the parametric method (\cite{zhang2010thesis}) and the Min-Max method (\cite{liu2011min}). Using all these methods,  we first estimated the optimal coefficient vector $\boldsymbol{\beta}$ and then calculated the maximized HUM values at those solutions. The above procedure was repeated for 500 times to obtain the mean and standard error of the optimal solutions of $\boldsymbol{\beta}$ and the corresponding HUM $D_{E}(\boldsymbol{\beta})$. The mean and standard errors of HUM for different methods are reported in Table \ref{tab-simu-hum}, whereas those values for the coefficient vector are reported in Table \ref{tab-simu-beta}.

Under all the above scenarios, the proposed SSHUM and NSHUM methods outperform the other existing approximation methods. Under the first three scenarios where biomarkers' values are generated from normal distributions, the SSHUM and NSHUM methods performs as good as the parametric method in Section 3.1 and outperform the Frechet bounds method and Min-Max method. In Scenario 4 where biomarkers' values are non-normally distributed, the parametric method with normality assumption performs poorly than the proposed methods. However, there is no observable difference in the accuracy measure between the SSHUM and NSHUM methods, suggesting both the sigmoid and the normal CDF approximations perform equally good for non-normal distributions.

\begin{table}[ht]
\caption{Means and standard errors (in parenthesis) of obtained EHUM values at the optimal coefficient vector estimated using the methods: the Empirical method (\cite{zhang2011li}), the Fr\'echet bounds method (\cite{hsu2016chen}), the parametric method (\cite{zhang2010thesis}), the Min-Max method, SSHUM and NSHUM for simulation Scenarios 1, 2, 3, 4 with sample sizes $(60,60,60), (90,90,90), (120,120,120)$, based on 1000 repetitions. }
\centering
\scalebox{0.8}{
\begin{tabular}{l c c c c c c}
  \hline  \hline 
 ($n_1$, $n_2$, $n_3$) &  Empirical & Min-Max & Parametric & Fr\'echet & SSHUM & NHSUM \\ \hline
   \multicolumn{7}{c}{{\bf Scenario} 1 (True HUM=0.833)}  \\ \hline
(60, 60, 60)     & 0.824 (0.032)  & 0.804 (0.035)  & 0.826 (0.032) & 0.813 (0.034)  & 0.828  (0.033)  & 0.828 (0.033) \\
(90, 90, 90)      & 0.825 (0.026)  & 0.805 (0.028)  & 0.827 (0.027) & 0.815 (0.026)  & 0.827  (0.026)  & 0.827 (0.026) \\
(120, 120, 120)  & 0.824 (0.022)  & 0.804 (0.023)  & 0.825 (0.022) & 0.813 (0.022)  & 0.825  (0.022)  & 0.825 (0.022) \\ \hline
   \multicolumn{7}{c}{{\bf Scenario} 2 (True HUM=0.720)}  \\ \hline   
(60, 60, 60)    & 0.747 (0.039) & 0.734 (0.039) & 0.752 (0.039) & 0.744 (0.039) & 0.754 (0.039) & 0.754 (0.039)  \\
(90, 90, 90)     & 0.748 (0.032) & 0.735 (0.032) & 0.750 (0.031) & 0.744 (0.032) & 0.752 (0.031) & 0.752 (0.031)  \\
(120, 120, 120) & 0.749 (0.026) & 0.736 (0.027) & 0.751 (0.026) & 0.745 (0.027) & 0.752 (0.026) & 0.752 (0.026)  \\ \hline   
 \multicolumn{7}{c}{{\bf Scenario} 3 (True HUM=0.770)}  \\ \hline   
(60, 60, 60)    & 0.766 (0.037) & 0.752 (0.038) & 0.770 (0.037) & 0.756 (0.039) & 0.773 (0.037) & 0.773 (0.037)  \\
(90, 90, 90)     & 0.767 (0.030) & 0.753 (0.031) & 0.769 (0.031) & 0.756 (0.031) & 0.771 (0.030) & 0.771 (0.030)  \\
(120, 120, 120) & 0.769 (0.026) & 0.754 (0.026) & 0.770 (0.026) & 0.758 (0.026) & 0.771 (0.026) & 0.772 (0.026)  \\ \hline
\multicolumn{7}{c}{{\bf Scenario} 4 (True HUM=0.514)}  \\ \hline   
(60, 60, 60)    & 0.452 (0.059) & 0.412 (0.044) & 0.436 (0.057) & 0.391 (0.043) & 0.521 (0.045) & 0.521 (0.046)  \\
(90, 90, 90)     & 0.474 (0.051) & 0.412 (0.036) & 0.425 (0.058) & 0.391 (0.038) & 0.515 (0.036) & 0.515 (0.036)  \\
(120, 120, 120) & 0.484 (0.046) & 0.411 (0.031) & 0.420 (0.047) & 0.392 (0.033) & 0.512 (0.031) & 0.512 (0.031)  \\ 
 \hline 
 \hline 
\end{tabular}\label{tab-simu-hum}
}
\end{table}

\begin{table}[htb]
\caption{Means (biases and standard errors) of $(\beta_{1}, \beta_{2})^{T}$ (based on 1000 replications) by different methods for {Scenario} 1. All the methods were maximized using Quasi-Newton method.}
\centering
\scalebox{0.65}{
\begin{tabular}{lc cc c c c }
   \hline  \hline  
     Sample size   & $(\beta_{1}, \beta_{2})^{T}$ & Empirical & Parametric & Fr\'echet & SSHUM & NSHUM \\                
  \hline   
      \multicolumn{7}{c}{{\bf Scenario} 1}  \\ \hline   
  $n=(60,60,60)$    & 1.2 & 1.045 (-0.155, 0.179) & 1.230 (0.030, 0.294) & 1.995 (0.795, 0.067) & 1.275  (0.075, 0.367) & 1.308 (0.108, 0.377) \\
                    & 1.1 & 1.018 (-0.082, 0.170) & 1.124 (0.024, 0.284) & 1.990 (0.890, 0.070) & 1.182  (0.082, 0.382) & 1.215 (0.115, 0.384) \\                     
                     \hline      
  $n=(90,90,90)$    & 1.2 & 1.050 (-0.15, 0.125) &1.230 (0.030, 0.229) &1.998 (0.798, 0.062) &1.274 (0.074, 0.297) &1.282 (0.082, 0.311)\\
                    & 1.1 & 1.010 (-0.09, 0.113) &1.125 (0.025, 0.219) &1.990 (0.890, 0.062) &1.175 (0.075, 0.289) &1.178 (0.078, 0.299)\\
                     \hline                   
  $n=(120,120,120)$ & 1.2 &1.074 (-0.126, 0.135) &1.219 (0.019, 0.200) &1.994 (0.794, 0.059)  &1.256 (0.056, 0.238) &1.258 (0.058, 0.246)\\
                    & 1.1 &1.013 (-0.087, 0.114) &1.117 (0.017, 0.184) &1.973 (0.873, 0.092)  &1.144 (0.044, 0.215) &1.148 (0.048, 0.224)\\                                                         
   \hline \hline 
   \multicolumn{7}{c}{{\bf Scenario} 2}  \\ \hline   
   $n=(60,60,60)$   & 1.378 &1.059 (-0.320, 0.180) &1.502 (0.124, 0.557) &2.000 (0.622, 0.066) &1.628 (0.25, 0.657) &1.670 (0.292, 0.658) \\
                    & 1.189 &1.006 (-0.183, 0.139) &1.291 (0.102, 0.503) &1.994 (0.805, 0.068) &1.399 (0.21, 0.616) &1.446 (0.257, 0.598) \\                     
                     \hline       
  $n=(90,90,90)$    & 1.378 &1.086 (-0.293, 0.218) &1.457 (0.079, 0.447) &2.003 (0.625, 0.087) &1.546 (0.168, 0.549) &1.577 (0.198, 0.526) \\
                    & 1.189 &1.019 (-0.170, 0.174) &1.259 (0.070, 0.395) &1.986 (0.797, 0.081) &1.337 (0.148, 0.484) &1.369 (0.180, 0.479) \\
                     \hline                   
  $n=(120,120,120)$ & 1.378 &1.111 (-0.267, 0.237) &1.414 (0.036, 0.338) &2.006 (0.628, 0.102) &1.474 (0.096, 0.411) &1.485 (0.107, 0.409) \\
                    & 1.189 &1.025 (-0.164, 0.185) &1.216 (0.027, 0.307) &1.977 (0.788, 0.132) &1.272 (0.082, 0.381) &1.282 (0.093, 0.382) \\                                                         
   \hline \hline 
    \multicolumn{7}{c}{{\bf Scenario} 3}  \\ \hline   
   $n=(60,60,60)$   & 1.256 &1.058 (-0.199,   0.167) & 1.299 (0.042, 0.345) &2.011 (0.754, 0.246) &1.400 (0.144, 0.490) &1.446 (0.189, 0.515) \\
                    & 0.903 &0.964 ( 0.062,   0.137) & 0.947 (0.044, 0.323) &1.983 (1.081, 0.068) &1.032 (0.130, 0.435) &1.086 (0.183, 0.471) \\                     
                     \hline       
  $n=(90,90,90)$    & 1.256 &1.102 (-0.154, 0.255) & 1.292 (0.036, 0.284) &2.003 (0.746, 0.073) &1.338 (0.082, 0.350)  &1.362 (0.106, 0.370) \\
                    & 0.903 &0.957 ( 0.054, 0.206) & 0.932 (0.029, 0.253) &1.977 (1.075, 0.086) &0.974 (0.071, 0.318)  &0.993 (0.091, 0.346) \\
                     \hline                  
  $n=(120,120,120)$ & 1.256 &1.122 (-0.135, 0.189) & 1.284 (0.028, 0.240)  &2.005 (0.748, 0.089) &1.324  (0.067, 0.301) &1.328 (0.071, 0.319) \\
                    & 0.903 &0.940 ( 0.038, 0.144) & 0.917 (0.015, 0.224)  &1.965 (1.062, 0.105) &0.948  (0.045, 0.277) &0.951 (0.048, 0.291) \\
                    \hline \hline 
                    \multicolumn{7}{c}{{\bf Scenario} 4}  \\ \hline   
   $n=(60,60,60)$   & 0.047 & 0.695 (0.648, 0.368) & 0.964 (0.917, 0.927) & 1.960 (1.913,   0.233)  & 0.089 (0.042, 0.091) & 0.100 (0.053, 0.130) \\
                    & 0.456 & 1.028 (0.571, 0.538) & 3.237 (2.781, 4.707) & 3.894 (3.437,  41.735)  & 0.530 (0.074, 0.225) & 0.563 (0.107, 0.276) \\                     
                     \hline       
  $n=(90,90,90)$    & 0.047 & 0.440 (0.393, 0.387) & 1.031  (0.984, 0.817) & 1.641 (1.594, 8.895)  & 0.079 (0.032, 0.055) & 0.080 (0.033, 0.061) \\
                    & 0.456 & 0.925 (0.469, 0.359) & 2.450  (1.993, 1.572) & 2.324 (1.868, 8.906)  & 0.505 (0.049, 0.159) & 0.513 (0.056, 0.171) \\
                     \hline                  
  $n=(120,120,120)$ & 0.047 & 0.306  (0.259, 0.345)  & 1.173 (1.126, 1.059) & 1.888 (1.841, 0.308) & 0.073 (0.025, 0.046) & 0.073 (0.026, 0.046) \\
                    & 0.456 & 0.838  (0.382, 0.344)  & 2.548 (2.091, 1.893) & 2.020 (1.564, 0.406) & 0.492 (0.036, 0.140) & 0.494 (0.037, 0.144) \\  
   \hline \hline  
\end{tabular}\label{tab-simu-beta}
 }
\end{table}

%
%
%

\section{Real Data Analysis}

\subsection{{The Alzheimer's  Disease Data Analysis}}

The first data set that we analyzed here for illustration is a subset of the longitudinal cohort data on Alzheimer's Disease (AD) from Alzheimer's Disease Research Center (ADRC) at Washington University. The dataset is available in the R package \texttt{DiagTest3Grp} ({\it https://www.cran.org}). In this data set, measurements of 14 neuro-psychological markers  were collected from 118 independent individuals of age 75 among which 44 individuals were non-demented, 43 were very mildly demented, and 21 individuals were mildly demented. It is now commonly accepted that treatment for Alzheimer's disease is a rather complicated issue and more clinically useful strategy is to apply appropriate interventions for earlier stage patients with relatively mild conditions (\cite{Dubois2014},\cite{Dubois2016}). Therefore it is meaningful to differentiate three or even more categories of patients with ascending disease severity and subsequently offer category-specific treatments.

Due to some missing observations, we deleted 10 individuals from the data set for our analysis. Note that values of these fourteen biomarkers can be negative. Furthermore, as we can see from the boxplot in Figure \ref{fig-boxplot-AL} and density plot in \ref{fig-density-AL}, there is a clear decreasing trend in the distributions of all the fourteen neuro-psychological markers across the dementia status. This shows the potential discrimination power of these individual markers. This observation was further evident by their individual discrimination power in terms of EHUM values where \textit{factor1}, \textit{ktemp} and \textit{zpsy004} have the highest individual EHUM values ranging from 0.70 to 0.78. Even the lowest EHUM values for the individual markers lie above 0.3, clearly much larger than the lowest EHUM value for random guess which is 0.17 in this case. 

To see the improvement in discrimination accuracy by combing these individual markers over the individual markers and to facilitate comparison, we employed all the six combining methods discussed in Section 3. The estimated EHUM values with their respective standard errors using all the six methods are reported in Table \ref{tab-AL3} along with the coefficient parameter estimates and their respective bootstrap standard errors. We note that the empirical method has the highest EHUM value of 0.832 which is a substantive improvement than the highest individual biomarker's EHUM value of 0.784. The SSHUM method has the second largest EHUM value of 0.828, also a substantive improvement over the individual biomarkers. However, as we can see the Min-Max and Naive method (where we assumed equal weight for each individual biomarkers) have the lowest EHUM values of 0.80 and 0.792, respectively. 

National Institute of Aging-Alzheimer's Association (NIA-AA) published research criteria for AD diagnosis in 2011 using biomarkers information. In addition to dementia due to AD, other stages of interest include prodromal AD (mild cognitive impairment) and preclinical AD (individuals with normal condition with AD pathology). The markers evaluated in our analysis may also offer useful insight for such mutli-stage diagnosis.

 \begin{figure}[ht]
 \centering
  \includegraphics[scale=0.5]{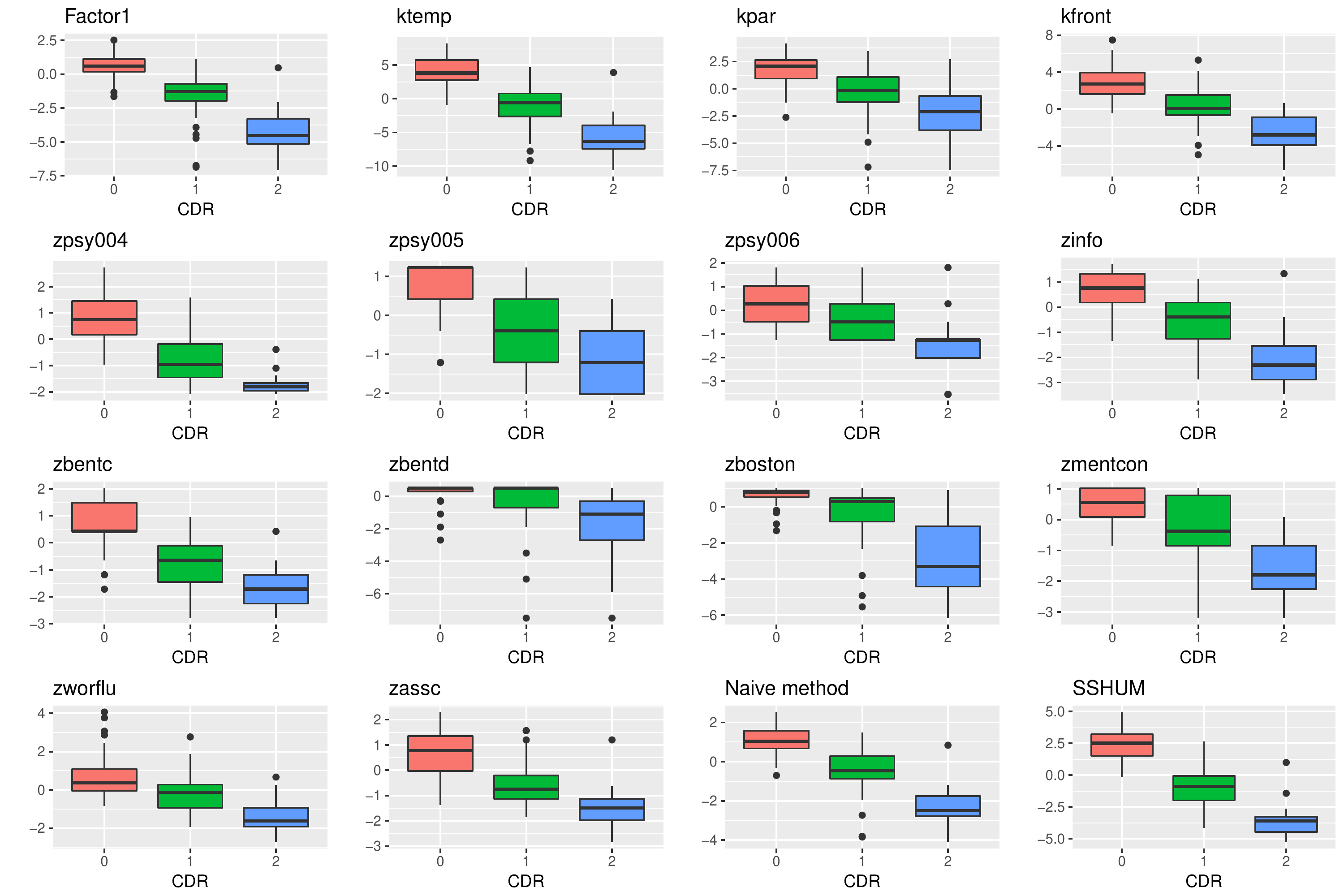} 
 \caption{Boxplot for individual and combined biomarkers for Alzheimer data set.}
 \label{fig-boxplot-AL}
\end{figure}

 \begin{figure}[ht]
 \centering
  \includegraphics[scale=0.5]{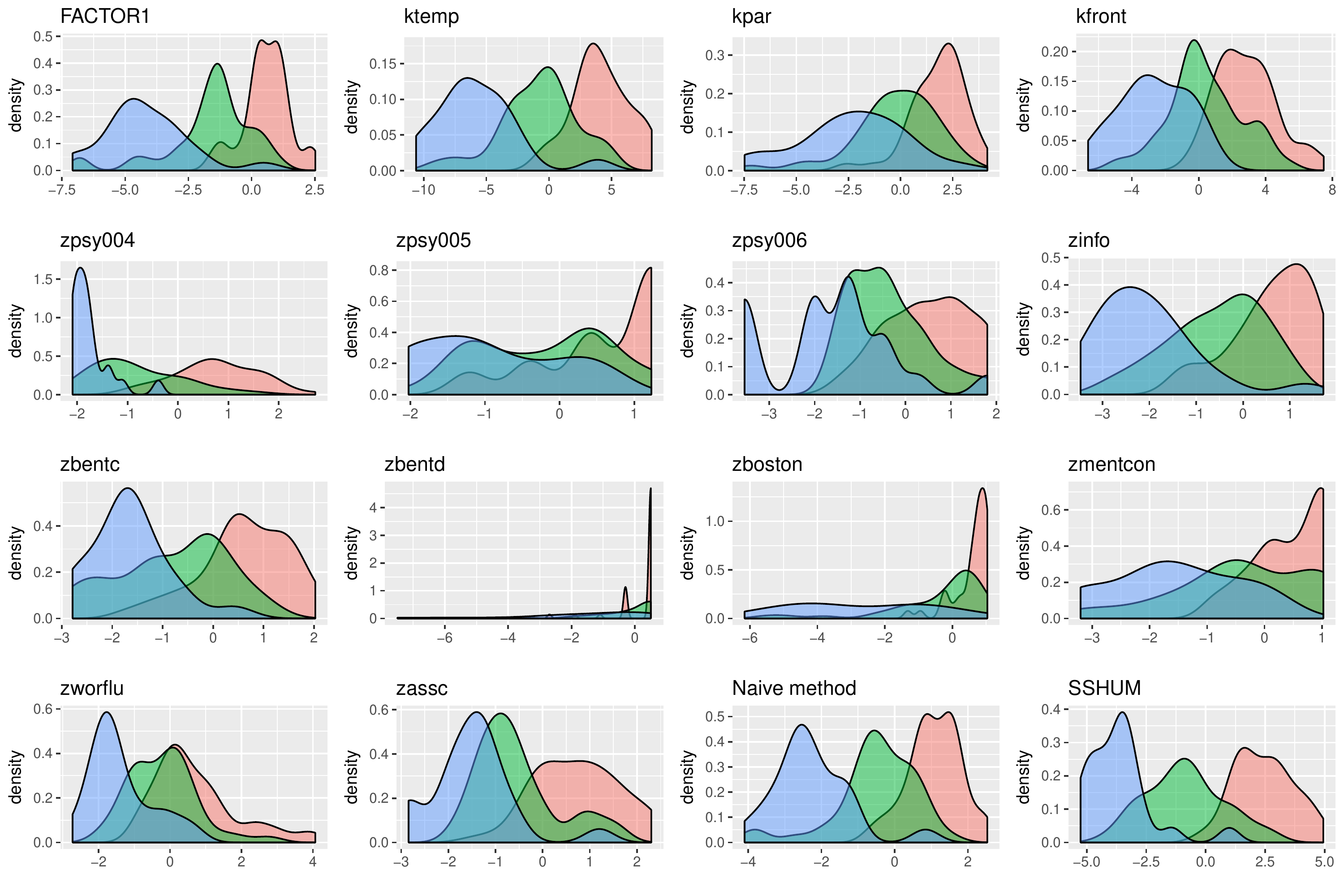} 
 \caption{Density plot for individual and combined biomarkers for Alzheimer data set.}
 \label{fig-density-AL}
\end{figure}

\begin{table}[ht]
\caption{Empirical HUM values (with bootstrap standard errors) for the individual biomarkers for the AKI and Alzheimer data sets.}
\centering
\scalebox{0.7}{
\begin{tabular}{cc | cc}
  \hline 
   \multicolumn{2}{c}{Alzheimer data} & \multicolumn{2}{c}{ERICCA data} \\
  \hline
  Individual biomarkers & HUM (se) & Individual biomarkers & HUM (se)  \\
  \hline
 FACTOR1   & 0.774 (0.056)  &  NGAL 0 hours  & 0.179 (0.029)  \\
 ktemp     & 0.784 (0.055)  &  NGAL 6 hours  & 0.222 (0.034)  \\
 kpar      & 0.600 (0.065)  &  NGAL 12 hours & 0.273 (0.040)  \\
 kfront    & 0.654 (0.059)  &  NGAL 24 hours & 0.315 (0.042)  \\
 zpsy004   & 0.718 (0.058)  &                &                \\
 zpsy005   & 0.316 (0.064)  &                &                \\ 
 zpsy006   & 0.442 (0.069)  &                &                \\
 zinfo     & 0.643 (0.065)  &                &                \\
 zbentc    & 0.506 (0.060)  &                &                \\
 zbentd    & 0.144 (0.047)  &                &                \\
 zboston   & 0.590 (0.066)  &                &                \\
 zmentcon  & 0.367 (0.065)  &                &                \\
 zworflu   & 0.561 (0.066)  &                &                \\            
 zassc     & 0.648 (0.066)  &                &                \\
\hline 
\end{tabular}\label{tab-AL1}
}
\end{table}

\begin{table}[htb]
\caption{{Estimated coefficients and the HUM values (with standard errors in parenthesis) for the Alzheimer's disease data using Naive method, Empirical, SSHUM, NSHUM, Fr\'echet, Parametric and Min-max methods based on 100 repetitions.}}
\centering
\scalebox{0.6}{
\begin{tabular}{c c c c c c c c}
  \hline  \hline
Biomarkers& $\beta_{Naive}$ & $\beta_{Empirical}$ & $\beta_{SSHUM}$ & $\beta_{NSHUM}$ & $\beta_{Frechet}$ & $\beta_{Parametric}$ & $\beta_{Min-Max}$ \\ \hline
FACTOR1   & 0.267 &   0.437 (0.2056) & -0.092 (0.2457) &  0.431 (0.2235) &  0.192 (0.2064) & -0.032 (0.2235) & - \\
ktemp     & 0.267 &   0.117 (0.1613) &  0.233 (0.1332) &  0.155 (0.1218) &  0.192 (0.1564) &  0.203 (0.1774) & - \\
kpar      & 0.267 &   0.154 (0.1379) &  0.261 (0.1278) &  0.229 (0.1389) &  0.161 (0.1526) &  0.019 (0.2000) & - \\
kfront    & 0.267 &  -0.005 (0.1590) & -0.006 (0.1577) & -0.030 (0.1628) &  0.343 (0.1713) &  0.232 (0.1860) & - \\
zpsy004   & 0.267 &   0.685 (0.1953) &  0.447 (0.0986) &  0.433 (0.0911) &  0.667 (0.2171) &  0.495 (0.2301) & - \\
zpsy005   & 0.267 &   0.173 (0.1604) &  0.063 (0.1683) &  0.071 (0.1381) &  0.192 (0.1909) &  0.082 (0.2191) & - \\
zpsy006   & 0.267 &   0.180 (0.1841) &  0.402 (0.1270) &  0.318 (0.1263) &  0.192 (0.1915) &  0.423 (0.2071) & - \\
zinfo     & 0.267 &  -0.283 (0.2664) & -0.447 (0.1730) & -0.433 (0.2389) & -0.073 (0.2699) & -0.244 (0.2657) & - \\
zbentc    & 0.267 &  -0.043 (0.1905) &  0.268 (0.1697) &  0.063 (0.1599) & -0.262 (0.2033) &  0.249 (0.2205) & - \\
zbentd    & 0.267 &   0.007 (0.2093) & -0.401 (0.2179) & -0.291 (0.2478) &  0.001 (0.2293) & -0.067 (0.2403) & - \\
zboston   & 0.267 &   0.173 (0.1983) & -0.128 (0.1641) &  0.183 (0.1662) &  0.000 (0.2045) &  0.303 (0.2149) & - \\
zmentcon  & 0.267 &   0.235 (0.2564) &  0.139 (0.1466) &  0.288 (0.1462) & -0.196 (0.2704) & -0.377 (0.2172) & - \\
zworflu   & 0.267 &  -0.192 (0.2175) & -0.138 (0.1899) &  0.021 (0.1797) & -0.065 (0.2387) &  0.326 (0.2352) & - \\
zassc     & 0.267 &  -0.189 (0.2580) & -0.125 (0.2503) & -0.222 (0.2379) &  0.384 (0.2838) & -0.079 (0.2395) & - \\
 \hline 
 HUM      & 0.792 & 0.832 (0.0545)  & 0.874 (0.0179) &  0.849 (0.0177) &  0.812 (0.0614) & 0.817 (0.0584) & 0.800 (0.0509) \\
 \hline
\end{tabular}\label{tab-AL3}
}
\end{table}

\subsection{The ERICCA data analysis}
Here we analyze an acute kidney injury dataset following a heart surgery to illustrate our proposed method. We consider the data from the {\bf E}ffect of {\bf R}emote {\bf I}schemic Preconditioning on {\bf C}linical Outcomes in Patient Undergoing {\bf C}oronary {\bf A}rtery Bypass Graft Surgery (ERICCA) trial where a group of 1612 patients participated in a cardiovascular surgery and were observed for one year after the surgery (\cite{hausenloy2012,hausenloy2015}). All the patients were randomized to two different methods of surgeries namely Remote Ischemic Conditioning (RIC) or Sham Preconditioning. During the study period, some patients developed a disease called Acute Kidney Injury (AKI) along with few other diseases post-surgery. The AKI was recorded as a multi-category ordinal outcome with four levels based on the severity level.  The data also includes cardiovascular death and all-cause mortality at 1 year (binary), non-fatal Myocardial Infarction (MI) (binary) and coronary revascularization or stroke at 1 year (binary). In literature, studies on prediction of AKI after cardiac surgery has been performed in several occasions. Assuming AKI as a binary outcome, \cite{heeraj2017} found that the serum Neutrophil Gelatinase Associated Lipocalin (NGAL) measurements taken at 0 (before surgery), 6, 12 and 24 hours after surgery are significant influential biomarkers in the development of AKI. In addition, they showed that for the risk-stratification of patients prior to cardiac surgery for AKI may be improved by adding pre-oprative levels of NGAL to existing risk scores where existing risk score was calculated based on age, gender, diabetes mellitus, hypertension, peripheral vascular disease, previous Coronary Artery Bypass Graft type of surgery planned, use of intra-aortic ballon pump and few other baseline covariates. However, the main limitation of their study is that they did not consider the multiple categories of the AKI outcome. Instead, they converted it to binary outcome where level 0 stands for no AKI and level 1 stands for any of the 1,2,3 levels of AKI in the data.

 To illustrate the proposed method, we consider the AKI within 72 hours of surgery as our multi-category outcome which are leveled as 0 (none), 1, 2, 3 as per the international Kidney Disease: Improving Global Outcomes
classification (KDIGO) criteria on serum creatinine. Since level 3 has only a few observations, we combine the levels 2 and 3 into a single category denoted as the highest risk group. Therefore, in the following analysis, the AKI has three levels/categories. Our biomarkers of interest in predicting AKI are individual NGAL at 0 (before surgery), 6, 12 and 24 hours after surgery and their different combinations using different methods.
In a previous analysis, \cite{heeraj2017} observed that there is a significant increase in AKI as the individual's pre-operative NGAL increases from the first to the third tertile ($>$220 ng/L). Hence they considered only the individuals from the third tertile and concluded that the pre-operative NGAL is a significant predictor in predicting binary AKI. There are 305 individuals  in our sample after discarding all the missing observations. Among these subjects, 172 patients did not develop AKI within the 72 hours of surgery (AKI=0), 99 patients developed level 1 AKI, and 34 developed level 2 (i.e., combined levels 2 and 3 in original scale) AKI.

{Note that larger values of the NGAL measurements indicate the higher level of severity of AKI. Since the NGAL measurements are highly skewed-distributed and large in number, so we transformed them into the logarithm scale to scale down those high numbers and make the distributions close to normal distributions. Considering logarithmic transformation of the biomarkers is a common strategy for this type of data analysis (see e.g., \cite{pepe2000thompson}). To see the visual discrimination power of these individual log of NGAL measurements, the box plots and the density plots are shown in Figures \ref{fig-boxplot-ericca} and \ref{fig-density-ericca}, respectively. The estimated empirical HUM values for the individual NGAL at four different time points are  0.179 (at 0 hours),  0.222 (at 6 hours), 0.273 (at 12 hours),  and 0.315 (at 24 hours). These values are also reported in Table \ref{tab-AL1}, along with their respective standard error. Recall that for random guess the HUM value is 1/6=0.1667 when the disease outcome variable has three possible outcomes. That is to say HUM value for any biomarker less that 0.1667 indicates that the biomarker is weaker in predicting the disease outcome and should be avoided from the prediction model. In this case, all the NGAL measurements can be included in the prediction model. Further, it is noticed that as the time of NGAL measurement increases from 0 hours to 24 hours, the HUM value  increases to almost two times that of the 0 hours. It indicates the strong discrimination power of the NGAL biomarker in predicting AKI as time progresses after surgery.  
}

 \begin{figure}[ht]
 \centering
  \includegraphics[scale=0.5]{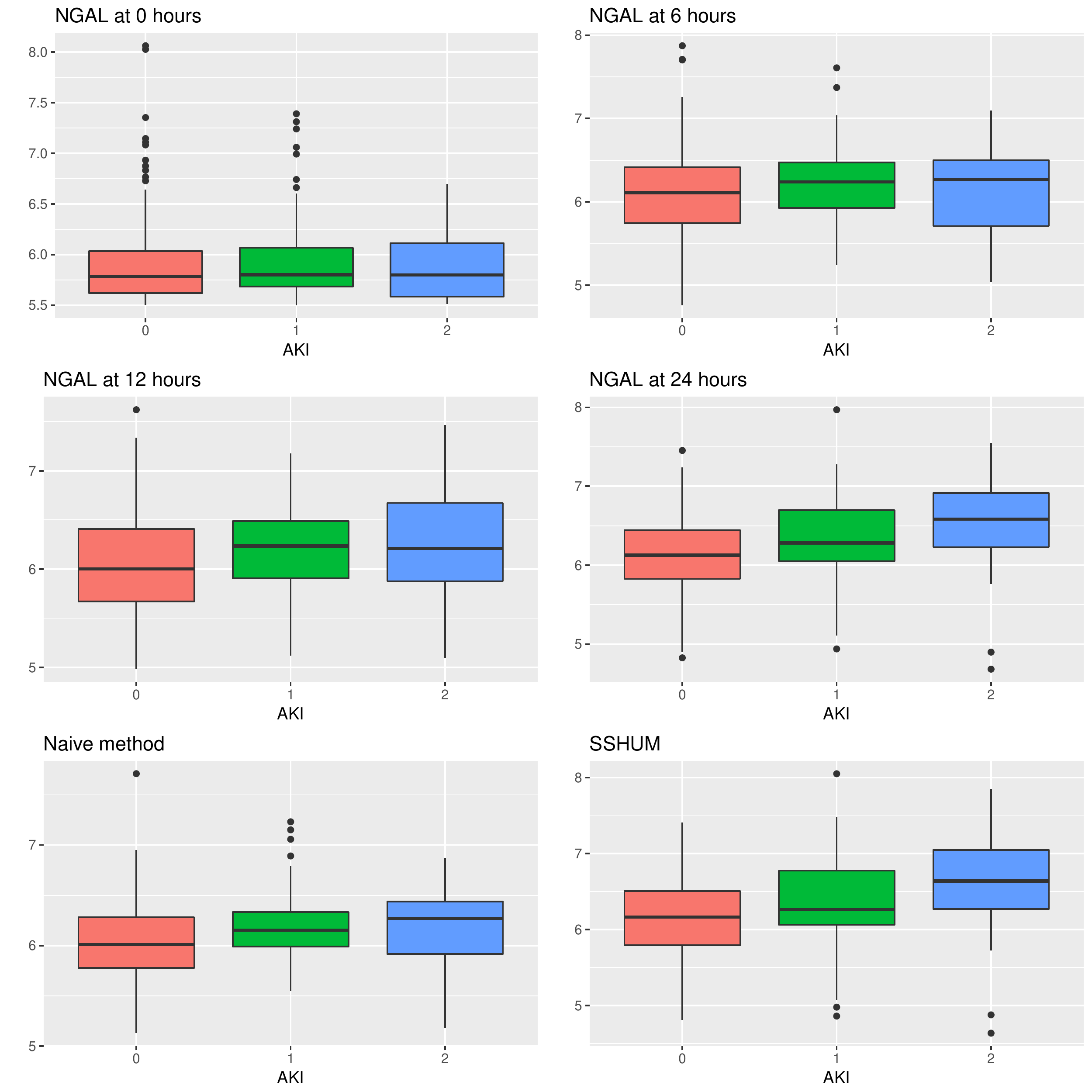} 
 \caption{{Boxplot for individual and combined NGALs for ERICCA data set. The top 4 plots represents the NGAL levels at 0, 6, 12 and 24 hours after the surgery for 3 levels of AKI. Bottom left diagram shows the boxplots for Naive method (i.e., linear combination of covariates with equal positive coefficients) and the bottom right diagram shows the boxplots for SSHUM method.}}
 \label{fig-boxplot-ericca}
\end{figure}

 \begin{figure}[ht]
 \centering
  \includegraphics[scale=0.5]{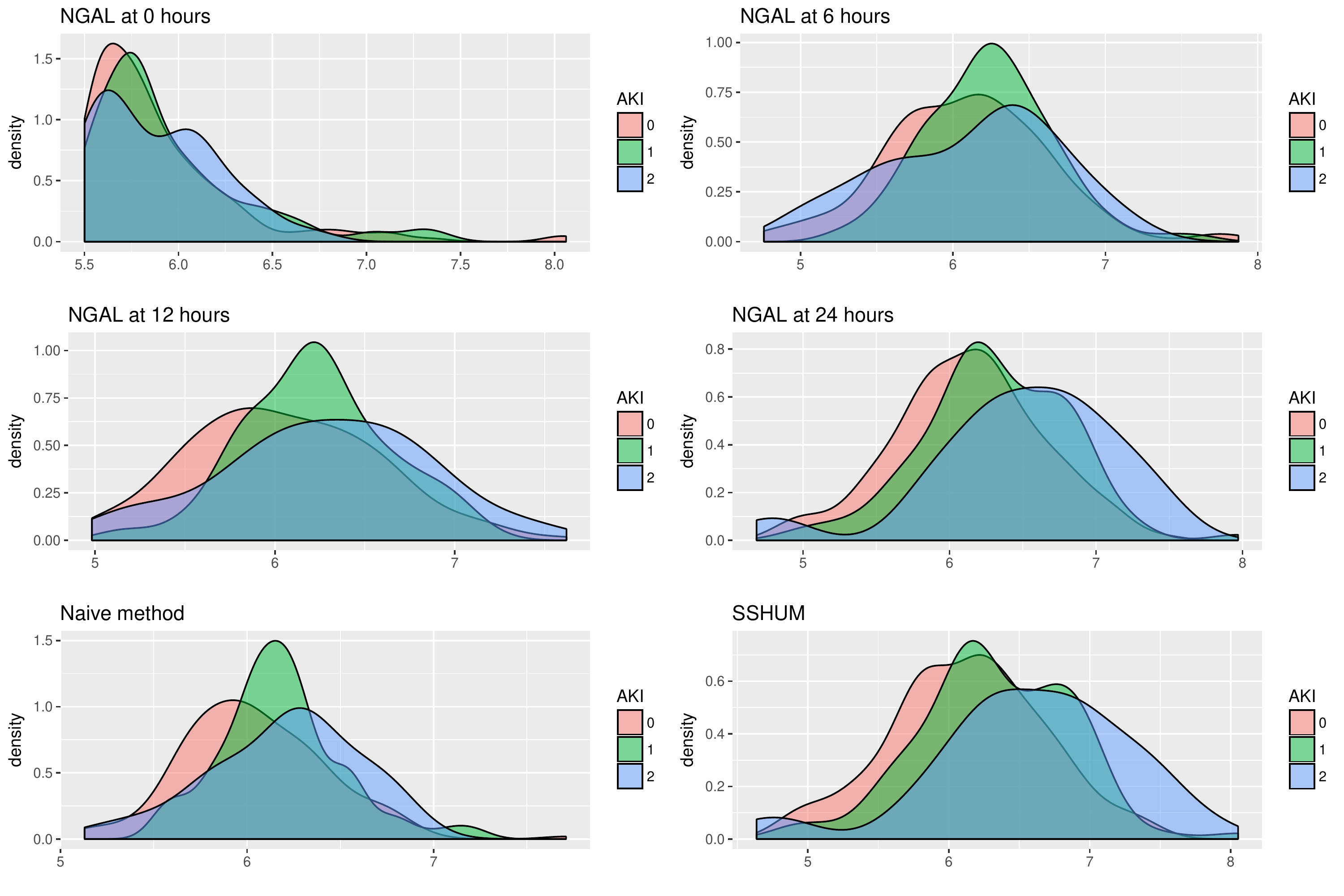} 
 \caption{Density plot for individual and combined NGALs for ERICCA data set.}
 \label{fig-density-ericca}
\end{figure}

 Further, we treat the four NGAL measurements as four biomarkers and apply our proposed SSHUM method to combine these markers. As comparison, a naive linear combination approach with equal weights on the four markers is also constructed. The distributions of these combined markers are also displayed in  Figures \ref{fig-boxplot-ericca} and \ref{fig-density-ericca}. It is noted that SSHUM separates the three class in the most effective way. 

 Further, we obtain the HUM values for other existing methods along with their respective optimal linear combination estimates. The estimates along with their bootstrap standard errors are reported  in Table \ref{tab-AL2}. We note that all the linear combining methods yield larger HUM values than that of the individual biomarkers and the naive equal weight method. The proposed sigmoid approximation yields the highest HUM value compared to the other existing methods. Although the proposed method combines the time-varying NGAL measurements in a more effective way than the others, still further studies may be required to support the effectiveness of such NGAL measurements and their combining factor in predicting AKI. 
 

  \begin{table}[htb]
\caption{{Estimated optimal coefficients and the HUM values (with standard errors in parenthesis) for the ERICCA dataset using naive method, empirical method, SSHUM, NSHUM, Fr\'echet, parametric and Min-Max methods based on 100 repetitions.}}
\centering
\scalebox{0.6}{
\begin{tabular}{c c c c c c c c}
  \hline  \hline
Biomarkers& ${naive}$ & ${Empirical}$ & ${SSHUM}$ & ${NSHUM}$ & ${Frechet}$ & ${Parametric}$ & ${Min-Max}$  \\ \hline
NGAL 0 hours    & 0.5 &  0.412 (0.2869) & -0.208 (0.3078) & -0.097 (0.3142) & 0.234 (0.0798)  & 0.236 (0.3636) & -\\
NGAL 6 hours    & 0.5 & -0.050 (0.4074) & -0.660 (0.3201) & -0.387 (0.3196) & -0.382 (0.1083) & 0.593 (0.3659) & - \\
NGAL 12 hours   & 0.5 &  0.594 (0.2098) &  0.360 (0.3320) &  0.176 (0.3377) &  0.566 (0.0426) & 0.590 (0.2563) & - \\
NGAL 24 hours   & 0.5 &  0.688 (0.1917) &  1.508 (0.2570) &  0.900 (0.2665) &  0.692 (0.0462) & 0.494 (0.1762) & - \\
 \hline 
 HUM      & 0.281 & 0.317 (0.0154)  & 0.326 (0.0140) &  0.325 (0.0135) &  0.312 (0.0054) & 0.287 (0.0182) & 0.303 (0.0079) \\
 \hline
\end{tabular}\label{tab-AL2}
}
\end{table}

\section{Discussion} 
Improving diagnostic accuracy by combining multiple biomarkers have been studied both for binary and multi-category outcomes. In this article, we have extended the idea of direct maximization of empirical hyper-volume under manifolds, specifically volume under surface (VUS) proposed by \cite{zhang2011li}, to a smoothing approximation of it using a class of smooth CDFs which is controlled by a tuning parameter. In particular, we have used the logistic CDF (sigmoid function) and normal CDF to operationalize our proposed method. We have also discussed about the choice of the tuning parameter. Consistency and asymptotic normality of the coefficient estimators using the proposed method have been established. Furthermore, through simulation studies we observe that the proposed method is computationally less challenging than the direct maximization of the EHUM, which is non-smooth and non-differentiable. We also note that the performance of the proposed method heavily depends on the choice of the tuning parameter $\lambda$, with lower values of $\lambda$ leading to results very similar to the empirical method with less bias but large variability. This is a problem of bias-variance trade-off which we have discussed in considerable detail in Section 2.3. Results from our simulation study and the two real medical data analyses have shown that shown that in general, the proposed method outperforms other methods including the empirical method. To obtain the estimated coefficient vectors maximizing SHUM, we considered the step-down algorithm. However, in future, coming up with advanced computational aids and fast global optimization algorithms for simultaneous 
estimation of the whole coefficient vector (instead of estimating one at a time using step-down algorithm) maximizing
 SHUM might further improve the solutions.



\section*{Acknowledgements}
We thank Jon Wellner, Palash Ghosh and Heerajnarain Bulluck for helpful discussions. The work was partially supported by grants R-155-000-205-114, R-155-000-195-114, R-155-000-197-112, R-155-000-197-113 and MOE2015-T2-2-056 from the Ministry of Education in Singapore, as well as the start-up grant of Bibhas Chakraborty from Duke-NUS Medical School.

\bibliographystyle{wileyj}
\bibliography{references}

\section*{Appendix}

\subsection*{A1: Proof of Theorem 1}
Assuming (A1)-(A3), \cite{zhang2011li} proved the consistency of $\widehat{\boldsymbol{\beta}}_{E}$, an empirical HUM based estimator of ${\boldsymbol{\beta}}$  for three-category ordinal outcome, using the result of maximum rank correlation type estimators by \cite{han1987} . In fact, it can be shown that $\widehat{\boldsymbol{\beta}}_{E}$ is consistent estimator of ${\boldsymbol{\beta}}$ for any number of categories. The above result is equivalent to $$\sup_{\boldsymbol{\beta} \in B} \abs{{D}_{E}(\boldsymbol{\beta}) - D(\boldsymbol{\beta})}=o_{p}(1),$$ 
 i.e., $\sup_{\boldsymbol{\beta} \in B} \abs{{D}_{E}(\boldsymbol{\beta}) - D(\boldsymbol{\beta})}$ converges to $0$
 in probability.
 
 Similarly, to prove the probability convergence of 
 $\widehat{\boldsymbol{\beta}}_{s_{n}}$, the proposed SSHUM based estimator, we have to show that 
 $$\sup_{\boldsymbol{\beta} \in B} \abs{{D}_{s_{n}}(\boldsymbol{\beta}) - D(\boldsymbol{\beta})} = o_{p}(1).$$   
Note that, using the triangular inequality, we can write
 \begin{eqnarray}\label{thm1-eqn1}
  \sup_{\boldsymbol{\beta} \in B}\abs{{D}_{s_n}(\boldsymbol{\beta}) - D(\boldsymbol{\beta})} &=& \sup_{\boldsymbol{\beta} \in B} \abs{{D}_{s_n}(\boldsymbol{\beta}) - {D}_{E}(\boldsymbol{\beta}) + {D}_{E}(\boldsymbol{\beta}) - D(\boldsymbol{\beta})}, \nonumber \\
  &\le& \sup_{\boldsymbol{\beta} \in B} \abs{{D}_{s_n}(\boldsymbol{\beta}) -{D}_{E}(\boldsymbol{\beta})} + \sup_{\boldsymbol{\beta} \in B} \abs{{D}_{E}(\boldsymbol{\beta}) - D(\boldsymbol{\beta})} \nonumber \\
  &=& \sup_{\boldsymbol{\beta} \in B} \abs{{D}_{s_n}(\boldsymbol{\beta}) - {D}_{E}(\boldsymbol{\beta})} + o_{p}(1).
 \end{eqnarray}
 Hence, to prove the consistency of $\widehat{\boldsymbol{\beta}}_{s_n}$, it is sufficient to prove the following lemma.
 \begin{lemma} 
 Under the assumptions (A1)-(A3), 
 $$\sup_{\boldsymbol{\beta} \in B} \abs{{D}_{E}(\boldsymbol{\beta}) - D_{s_n}(\boldsymbol{\beta})} \overset{p}{\longrightarrow} 0$$
 as $n \rightarrow \infty$.
\end{lemma}

\subsection*{Proof of Lemma 1} 
For binary outcome, \cite{ma2007huang} proved the consistency of $\boldsymbol{\beta}_{s_{n}}$ by showing that 
$$\sup_{\boldsymbol{\beta} \in B}\abs{{D}_{s_n}(\boldsymbol{\beta}) - {D}_{E}(\boldsymbol{\beta})} = o_{p}(1).$$
Here, we use the same idea to prove that $\sup_{\boldsymbol{\beta} \in B}\abs{{D}_{s_n}(\boldsymbol{\beta}) - {D}_{E}(\boldsymbol{\beta})} = o_{p}(1)$ for multi-category ordinal outcome. Define an equivalent definition of $D_{E}(\boldsymbol{\beta})$ as 
\begin{eqnarray*}
 D_{E}(\boldsymbol{\beta}) &=& C \displaystyle\sum_{i_1\ne i_2 \ne \cdots \ne i_{M}} I(Y_{Mi_M} > \cdots >Y_{1i_1}) I(\boldsymbol{\beta}^{T}\mathbf{Z}_{i_{M}i_{M-1}}>0) 
I(\boldsymbol{\beta}^{T}\mathbf{Z}_{i_{M-1}i_{M-2}}>0)\cdots I(\boldsymbol{\beta}^{T}\mathbf{Z}_{i_{2}i_{1}}>0),
\end{eqnarray*}
where $C = \dfrac{1}{n(n-1)\cdots (n-M+1)}$, $\mathbf{Z}_{i_{j+1}i_{j}} = \mathbf{X}_{(j+1)i_{j+1}} - \mathbf{X}_{ji_{j}}$, and $Y_{ji_{j}}$, $j=1,2,\cdots, M$ are defined as $Y_{ji_{j}}=j$ if the $i_{j}$-th observation belongs to the $j$-th category, otherwise 0.

Similarly, we define an equivalent definition of SSHUM as
\begin{eqnarray*}
 D_{s_n}(\boldsymbol{\beta}) &=& C \displaystyle\sum_{i_1\ne i_2 \ne \cdots \ne i_{M}} I(Y_{Mi_M}  > \cdots >Y_{1i_1}) s_n(\boldsymbol{\beta}^{T}\mathbf{Z}_{i_{M}i_{M-1}}) 
 s_n(\boldsymbol{\beta}^{T}\mathbf{Z}_{i_{M-1}i_{M-2}})\cdots s_n(\boldsymbol{\beta}^{T}\mathbf{Z}_{i_{2},i_{1}})
\end{eqnarray*}
For any $\delta>0$, we can write
\begin{eqnarray*}
 \abs{D_{E}(\boldsymbol{\beta}) - D_{s_n}(\boldsymbol{\beta})} &\le & T_{n1} + T_{n2}
 \end{eqnarray*}
where 
\begin{eqnarray*}
 T_{n1} &=& C \displaystyle\sum_{i_1\ne i_2 \ne \cdots \ne i_{M}} I(Y_{Mi_M}  > \cdots >Y_{1i_1}) \\ 
  && \abs{I(\boldsymbol{\beta}^{T}\mathbf{Z}_{i_{M}i_{M-1}}>0) 
\cdots I(\boldsymbol{\beta}^{T}\mathbf{Z}_{i_{2}i_{1}}>0) -  s_n(\boldsymbol{\beta}^{T}\mathbf{Z}_{i_{M}i_{M-1}}) 
 \cdots s_n(\boldsymbol{\beta}^{T}\mathbf{Z}_{i_{2}i_{1}})}\\
 && I\left(\max_{1 \le j \le M-1}\abs{\boldsymbol{\beta}^{T}\mathbf{Z}_{i_{j+1}i_{j}}}\ge \delta\right)
\end{eqnarray*}
 and 
 \begin{eqnarray*}
   T_{n2} &=&  C \displaystyle\sum_{i_1\ne i_2 \ne \cdots \ne i_{M}} I(Y_{Mi_M}  > \cdots >Y_{1i_1}) \\ 
  && \abs{I(\boldsymbol{\beta}^{T}\mathbf{Z}_{i_{M}i_{M-1}}>0) 
\cdots I(\boldsymbol{\beta}^{T}\mathbf{Z}_{i_{2}i_{1}}>0) -  s_n(\boldsymbol{\beta}^{T}\mathbf{Z}_{i_{M}i_{M-1}}) 
 \cdots s_n(\boldsymbol{\beta}^{T}\mathbf{Z}_{i_{2}i_{1}})}\\
 && I\left(\max_{1 \le j \le M-1}\abs{\boldsymbol{\beta}^{T}\mathbf{Z}_{i_{j+1}i_{j}}} < \delta\right).
 \end{eqnarray*}
\cite{ma2007huang} showed that on the set $\{\abs{x} \ge \delta \}$, $\abs{s_{n}(x) - I(x>0)} \le \exp{(-\abs{x}/\sigma_{n})} < \exp{(-\delta/\sigma_{n})} \rightarrow 0$  uniformly as $\sigma_{n} \rightarrow 0$. Following this, it can be shown that 
$$s_{n}(x_1) \rightarrow I(x_1>0) \mbox{ uniformly on the set } \{\abs{x_1}\ge \delta\},$$
$$s_{n}(x_2) \rightarrow I(x_2>0) \mbox{ uniformly on the set } \{\abs{x_2}\ge \delta\},$$
$$\vdots$$
$$\mbox{and } s_{n}(x_{M-1}) \rightarrow I(x_{M-1}>0) \mbox{ uniformly on the set } \{\abs{x_{M-1}}\ge \delta\}.$$
It implies that on the set $\{\max_{1\le i \le M-1}\abs{x_i} \ge \delta\}$, $s_{n}(x_i) \rightarrow I(x_i>0)$ uniformly for all $i=1,2,\cdots,M-1$. Following this, we can write
\begin{eqnarray*}
 && \abs{s_{n}(x_1) s_{n}(x_2) \cdots s_{n}(x_{M-1}) - I(x_1>0)I(x_2>0) \cdots I(x_{M-1}>0)} \\
 &\le&  \abs{s_{n}(x_1) - I(x_1>0)} s_{n}(x_2) \cdots s_{n}(x_{M-1}) + \\
 && I(x_1>0) \abs{s_{n}(x_2) \cdots s_{n}(x_{M-1}) - I(x_2>0) \cdots I(x_{M-1}>0)}, \\
 &\le&  \abs{s_{n}(x_1) - I(x_1>0)} s_{n}(x_2) \cdots s_{n}(x_{M-1}) + \\
 && I(x_1>0) \abs{s_{n}(x_2) - I(x_2>0)} s_{n}(x_3) \cdots s_{n}(x_{M-1}) + \\
 && I(x_1>0) I(x_2>0) \abs{s_{n}(x_3) \cdots s_{n}(x_{M-1}) - I(x_3>0) \cdots I(x_{M-1}>0)},\\
 && \vdots\\
 &\le&  \abs{s_{n}(x_1) - I(x_1>0)} s_{n}(x_2) \cdots s_{n}(x_{M-1}) + \\
 && I(x_1>0) \abs{s_{n}(x_2) - I(x_2>0)} s_{n}(x_3) \cdots s_{n}(x_{M-1}) + \cdots +\\
 && I(x_1>0) I(x_2>0) \cdots I(x_{M-2}>0) \abs{s_{n}(x_{M-1}) - I(x_{M-1}>0)}, \\
 &=& o_{p}(1) + o_{p}(1) + \cdots + o_{p}(1) = o_{p}(1).
\end{eqnarray*}
Now replacing $x_j$ by $\boldsymbol{\beta}^{T}\mathbf{Z}_{i_{j+1},i_{j}}$ in the above derivation, we can see that $T_{n1}$ converges to 0 uniformly on set $B$. The second term can be bounded above as
$$T_{n2} \le C \displaystyle\sum_{i_1\ne i_2 \ne \cdots \ne i_{M}} I\left(\max_{1 \le j \le M-1} \abs{\boldsymbol{\beta}^{T}\mathbf{Z}_{i_{j+1}i_{j}}} < \delta\right).$$
Again by the uniform convergence of the U-process, the right hand side of the above equation converges to $P\left(\max_{1 \le j \le M-1} \abs{\boldsymbol{\beta}^{T}\mathbf{Z}_{i_{j+1}i_{j}}} < \delta\right)$ almost surely on $B$. Further, using order statistic result, we can write 
\begin{eqnarray*}
 P\left(\max_{1 \le j \le M-1} \abs{\boldsymbol{\beta}^{T}\mathbf{Z}_{i_{j+1}i_{j}}}< \delta\right) &=& P\left(\abs{\boldsymbol{\beta}^{T}\mathbf{Z}_{i_{M}i_{M-1}}}< \delta, \abs{\boldsymbol{\beta}^{T}\mathbf{Z}_{i_{M-1}i_{M-2}}}< \delta, \cdots, \abs{\boldsymbol{\beta}^{T}\mathbf{Z}_{i_{2}i_{1}}}< \delta\right)\\
 &\le& P\left(\abs{\boldsymbol{\beta}^{T}\mathbf{Z}_{i_{j+1}i_{j}}}< \delta\right)
\end{eqnarray*}
for all $j=1,2,\cdots,M-1$ over $B$. 
Under the assumptions (A2) and (A3), it can be shown that $P\left(\abs{\boldsymbol{\beta}^{T}\mathbf{Z}_{i_{j+1}i_{j}}}< \delta\right)$ converges to 0 uniformly over $B$ as $\delta$ goes to 0. Hence, it proves that  $\sup_{\boldsymbol{\beta} \in B}\abs{{D}_{s_n}(\boldsymbol{\beta}) - {D}_{E}(\boldsymbol{\beta})} = o_{p}(1)$. 


\subsection*{A2: Proof of Theorem 2} 

For simplicity, we denote $\boldsymbol{\beta}(\boldsymbol{\theta}) = \boldsymbol{\beta}$ and $\boldsymbol{\beta}(\widehat{\boldsymbol{\theta}}) = \widehat{\boldsymbol{\beta}}.$
Note that 
$$\widehat{\boldsymbol{\beta}}_{s_n} = \arg \max_{\boldsymbol{\theta}} D_{s_n}(\boldsymbol{\beta}).$$

Define
\begin{eqnarray*}
 \mathbf{G}_{n}(\boldsymbol{\beta}) &=& \dfrac{\partial}{\partial\boldsymbol{\theta}}D_{s_n}(\boldsymbol{\beta}) \\
 &=& \dfrac{1}{\displaystyle\prod_{j=1}^{M}n_{j}} \displaystyle\sum_{i_{1}=1}^{n_{1}}\displaystyle\sum_{i_{2}=1}^{n_{2}} \cdots \displaystyle\sum_{i_{M}=1}^{n_{M}} \dfrac{\partial}{\partial\boldsymbol{\theta}} \left[s_{n}(\boldsymbol{\beta}^{T}(\mathbf{X}_{Mi_{M}}-\mathbf{X}_{(M-1)i_{(M-1)}})) \cdots s_{n}(\boldsymbol{\beta}^{T}(\mathbf{X}_{2i_{2}}-\mathbf{X}_{1i_{1}})) \right] \\
 &=& \dfrac{1}{N} \displaystyle\sum_{i_{1}=1}^{n_{1}}\displaystyle\sum_{i_{2}=1}^{n_{2}} \cdots \displaystyle\sum_{i_{M}=1}^{n_{M}} \Psi(\mathbf{X}_{1i_{1}}, \mathbf{X}_{2i_{2}}, \cdots, \mathbf{X}_{Mi_{M}}; \boldsymbol{\beta})
\end{eqnarray*}
 where 
 \begin{eqnarray*}
 \Psi(\mathbf{X}_{1i_{1}}, \mathbf{X}_{2i_{2}}, \cdots, \mathbf{X}_{Mi_{M}}; \boldsymbol{\beta}) &=& \dfrac{\partial}{\partial\boldsymbol{\theta}} \left[s_{n}(\boldsymbol{\beta}^{T}(\mathbf{X}_{Mi_{M}}-\mathbf{X}_{(M-1)i_{(M-1)}})) \cdots s_{n}(\boldsymbol{\beta}^{T}(\mathbf{X}_{2i_{2}}-\mathbf{X}_{1i_{1}})) \right], \\
  &=& \dfrac{\partial}{\partial\boldsymbol{\theta}} \left[\displaystyle\prod_{j=1}^{M-1} s_{n}(\boldsymbol{\beta}^{T}\mathbf{Z}_{i_{(j+1)}i_{j}})\right] \\
  &=& \displaystyle\sum_{l=1}^{M-1} \left[\displaystyle\prod_{j=1}^{M-1} s_{n}(\boldsymbol{\beta}^{T}\mathbf{Z}_{i_{(j+1)}i_{j}})\right]\left(1-s_{n}(\boldsymbol{\beta}^{T}\mathbf{Z}_{i_{(l+1)}i_{l}})\right) \mathbf{Z}_{i_{(l+1)}i_{l}}^{(-d)} \\
  &=& \kappa_{n}(\mathbf{X}_{1i_{1}}, \mathbf{X}_{2i_{2}}, \cdots, \mathbf{X}_{Mi_{M}};\boldsymbol{\beta}) \displaystyle\sum_{l=1}^{M-1}\left(1-s_{n}(\boldsymbol{\beta}^{T}\mathbf{Z}_{i_{(l+1)}i_{l}})\right) \mathbf{Z}_{i_{(l+1)}i_{l}}^{(-d)}
 \end{eqnarray*} 
 with $\mathbf{Z}_{i_{j+1}i_{j}} = \mathbf{X}_{(j+1)i_{j+1}} - \mathbf{X}_{ji_{j}}$, $\mathbf{Z}^{(-d)} = (Z_{1}, \cdots, Z_{d-1})^{T}$
 and $N=\displaystyle\prod_{j=1}^{M}n_{j}$.
 By definition of $\widehat{\boldsymbol{\beta}}_{s_n}$,
 \begin{eqnarray*}
  \mathbf{G}_{n}(\widehat{\boldsymbol{\beta}}_{s_n}) &=& \mathbf{0},
 \end{eqnarray*}
 and $\boldsymbol{\beta}_{0}$ is such that $$E(\Psi(\mathbf{X}_{1i_{1}}, \mathbf{X}_{2i_{2}}, \cdots, \mathbf{X}_{Mi_{M}}; \boldsymbol{\beta}_{0})) = \mathbf{0}.$$
Since $\mathbf{G}_{n}(\boldsymbol{\beta})$ is differentiable function, and $\sqrt{n}(\widehat{\boldsymbol{\theta}}_{s_n} - \boldsymbol{\theta}_{0})=o_{p}(1)$ (result from Theorem 1), hence using Taylor's series expansion we can write 
  \begin{eqnarray*}
  && \mathbf{0} = \mathbf{G}_{n}(\widehat{\boldsymbol{\beta}}_{s_n}) = \mathbf{G}_{n}(\boldsymbol{\beta}_{0}) +  \mathbf{G}^{'}_{n}(\boldsymbol{\beta}_{0}) (\widehat{\boldsymbol{\theta}}_{s_n} - \boldsymbol{\theta}_{0}) + \mathbf{R}_{n}
  \end{eqnarray*}
  where $\mathbf{G}^{'}_{n}(\boldsymbol{\beta}_{0}) = \dfrac{\partial}{\partial\boldsymbol{\theta}^{T}} \mathbf{G}_{n}(\boldsymbol{\beta})\mid_{\boldsymbol{\beta}=\boldsymbol{\beta}_{0}}$ is a $d\times d$ matrix. 
  

  Assuming (A4), we can write
    \begin{eqnarray}\label{eqn_slustky}
   \sqrt{n}(\widehat{\boldsymbol{\theta}}_{s_n} - \boldsymbol{\theta}_{0}) &=&  \left[-\mathbf{G}^{'}_{n}(\boldsymbol{\beta}_{0}) \right]^{-1} \sqrt{n}\mathbf{G}_{n}(\boldsymbol{\beta}_{0}) +  \left[\mathbf{G}^{'}_{n}(\boldsymbol{\beta}_{0}) \right]^{-1}\sqrt{n}\mathbf{R}_{n}. 
 \end{eqnarray}
 
 Note that following Theorem 1 where we have $(\widehat{\boldsymbol{\theta}}_{s_n} - \boldsymbol{\theta}_{0}) = \mathbf{o}_{p}(1)$, we can write
$$\sqrt{n}\mathbf{R}_{n} \overset{p}{\longrightarrow} \mathbf{0}.$$

 
 Following the large sample distribution of multivariate U-statistic (see \cite{kowalski2008}), it can be shown that  
  $$\sqrt{n}\mathbf{G}_{n}(\boldsymbol{\beta}_{0}) \overset{d}{\longrightarrow} N_{d-1}\left(\mathbf{0}, \mathbf{B}(\boldsymbol{\beta}_{0})\right)$$ where 
  \begin{eqnarray*}
   && \mathbf{B}(\boldsymbol{\beta}_{0}) = \displaystyle\sum_{m=1}^{M}\rho_{m}^{2}\Sigma_{\psi_{m}}, \\
   && \Sigma_{\psi_{m}} = Var(\tilde{\Psi}_{m1}(\mathbf{X}_{m1})),  \\
   && \tilde{\Psi}_{m1}(\mathbf{X}_{m1}) = E(\Psi(\mathbf{X}_{11}, \mathbf{X}_{21}, \cdots, \mathbf{X}_{M1})| \mathbf{X}_{m1}),  \\
    && \rho_{m}^{2} = \dfrac{n}{n_{m}}, \quad m=1,2,\cdots,M.
  \end{eqnarray*}

Similarly, using the weak law of large numbers, it can be shown that
 $$-\mathbf{G}^{'}_{n}(\boldsymbol{\beta}_{0}) = \dfrac{1}{N} \displaystyle\sum_{i_{1}=1}^{n_{1}}\displaystyle\sum_{i_{2}=1}^{n_{2}} \cdots \displaystyle\sum_{i_{M}=1}^{n_{M}} -\dfrac{\partial}{\partial\boldsymbol{\theta}^{T}}\Psi(\mathbf{X}_{i_{1}}, \mathbf{X}_{i_{2}}, \cdots, \mathbf{X}_{i_{M}}; \boldsymbol{\beta}) \overset{p}{\longrightarrow} \mathbf{A}(\boldsymbol{\beta}_{0})$$ 
 where 
 \begin{eqnarray*}\label{eqn-wlln-A}
 \mathbf{A}(\boldsymbol{\beta}_{0}) &=& E\left(-\dfrac{\partial}{\partial\boldsymbol{\theta}^{T}} \Psi(\mathbf{X}_{i_{1}}, \mathbf{X}_{i_{2}}, \cdots, \mathbf{X}_{i_{M}}; \boldsymbol{\beta})\mid_{\boldsymbol{\beta}=\boldsymbol{\beta}_{0}}\right).
 \end{eqnarray*}
 
Using Slustky's theorem in equation \eqref{eqn_slustky}, we can write 
$$ \sqrt{n}(\widehat{\boldsymbol{\theta}}_{n} - \boldsymbol{\theta}_{0}) \overset{d}{\longrightarrow} N_{d-1}\left(\mathbf{0}, \Sigma(\boldsymbol{\beta}_{0}) \right)$$
where $\Sigma(\boldsymbol{\beta}_{0}) = \mathbf{A}(\boldsymbol{\beta}_{0})^{-1} \mathbf{B}(\boldsymbol{\beta}_{0}) [\mathbf{A}(\boldsymbol{\beta}_{0})^{-1}]^{T}$, known as sandwich variance formula.

Explicit form of the first derivative of $\Psi(\mathbf{X}_{i_{1}}, \mathbf{X}_{i_{2}}, \cdots, \mathbf{X}_{i_{M}}; \boldsymbol{\beta})$ is given as follows:
\begin{eqnarray*}
 \dfrac{\partial}{\partial\boldsymbol{\theta}^{T}} \Psi(\mathbf{X}_{i_{1}}, \mathbf{X}_{i_{2}}, \cdots, \mathbf{X}_{i_{M}}; \boldsymbol{\beta})
  &=& \dfrac{\partial^{2}}{\partial\boldsymbol{\theta} \partial\boldsymbol{\theta}^{T}} \left[\displaystyle\prod_{j=1}^{M-1} s_{n}(\boldsymbol{\beta}^{T}\mathbf{Z}_{i_{(j+1)}i_{j}})\right] \\
  &=& \left(\left( \dfrac{\partial^{2}}{\partial\theta_{u} \partial\theta_{v}} \left[\displaystyle\prod_{j=1}^{M-1} s_{n}(\boldsymbol{\beta}^{T}\mathbf{Z}_{i_{(j+1)}i_{j}})\right] \right)\right), \qquad u,v=1,2,\cdots,d-1, 
 \end{eqnarray*} 
where
\begin{eqnarray*}
  \dfrac{\partial^{2}}{\partial\theta_{u} \partial\theta_{v}} \left[\displaystyle\prod_{j=1}^{M-1} s_{n}(\boldsymbol{\beta}^{T}\mathbf{Z}_{i_{(j+1)}i_{j}}) \right] &=& \displaystyle\sum_{l=1}^{M-1} \kappa_{n}(\boldsymbol{\beta}) \delta_{n;v}(\boldsymbol{\beta}) \left(1- s_{n}(\boldsymbol{\beta}^{T}\mathbf{Z}_{i_{(l+1)}i_{l}}) \right) {Z}_{i_{(l+1)}i_{l};u} - \\
  & & \displaystyle\sum_{l=1}^{M-1} \kappa_{n}(\boldsymbol{\beta}) s_{n}(\boldsymbol{\beta}^{T}\mathbf{Z}_{i_{(l+1)}i_{l}}) \left(1- s_{n}(\boldsymbol{\beta}^{T}\mathbf{Z}_{i_{(l+1)}i_{l}}) \right) {Z}_{i_{(l+1)}i_{l};u} {Z}_{i_{(l+1)}i_{l};v},
\end{eqnarray*}
 $$\kappa_{n}(\boldsymbol{\beta}) = \displaystyle\prod_{j=1}^{M-1} s_{n}(\boldsymbol{\beta}^{T}\mathbf{Z}_{i_{(j+1)}i_{j}}) $$
and $$ \delta_{n;v}(\boldsymbol{\beta}) = \displaystyle\sum_{k=1}^{M-1} \left(1- s_{n}(\boldsymbol{\beta}^{T}\mathbf{Z}_{i_{(k+1)}i_{k}})\right) Z_{i_{(k+1)i_{k};v}}$$

\begin{eqnarray*}
 & & \Psi(\mathbf{X}_{i_{1}}, \mathbf{X}_{i_{2}}, \cdots, \mathbf{X}_{i_{M}}; \boldsymbol{\beta}) \Psi(\mathbf{X}_{i_{1}}, \mathbf{X}_{i_{2}}, \cdots, \mathbf{X}_{i_{M}}; \boldsymbol{\beta})^{T} \\
 &=&  \left[ \kappa_{n}(\boldsymbol{\beta}) \displaystyle\sum_{l=1}^{M-1}\left(1-s_{n}(\boldsymbol{\beta}^{T}\mathbf{Z}_{i_{(l+1)}i_{l}})\right) \mathbf{Z}_{i_{(l+1)}i_{l}}^{(-d)} \right] \left[ \kappa_{n}(\boldsymbol{\beta}) \displaystyle\sum_{l=1}^{M-1}\left(1-s_{n}(\boldsymbol{\beta}^{T}\mathbf{Z}_{i_{(l+1)}i_{l}})\right) \mathbf{Z}_{i_{(l+1)}i_{l}}^{(-d)} \right]^{T} \\
 &=& \kappa_{n}(\boldsymbol{\beta})^{2} \displaystyle\sum_{l=1}^{M-1} \displaystyle\sum_{k=1}^{M-1} \left(1-s_{n}(\boldsymbol{\beta}^{T}\mathbf{Z}_{i_{(l+1)}i_{l}})\right) \left(1-s_{n}(\boldsymbol{\beta}^{T}\mathbf{Z}_{i_{(k+1)}i_{k}})\right) \mathbf{Z}_{i_{(l+1)}i_{l}}^{(-d)} \mathbf{Z}_{i_{(k+1)}i_{k}}^{(-d)T}.
\end{eqnarray*}

%
%

\end{document}